\documentclass[twocolumn,aps,prl,superscriptaddress,floatfix]{revtex4-1}
\usepackage{graphicx,subfigure}
\usepackage{amsmath,amssymb,bm}
\usepackage{mathtools}
\usepackage{multirow}
\usepackage{makecell}
\usepackage{textcomp}
\usepackage{float}
\usepackage{color}
\usepackage[normalem]{ulem}
\usepackage{dsfont}
\usepackage{mathrsfs}
\usepackage{braket}
\usepackage{transparent}
\usepackage{xcolor}
\usepackage{hhline}
\usepackage{graphicx}
\usepackage{pdfpages}
\usepackage[export]{adjustbox}
\usepackage{hyperref}
\usepackage{physics}

\makeatletter
\AtBeginDocument{\let\LS@rot\@undefined}
\makeatother

\newcommand{\figone}{
 \begin{figure*}[t]
    \centering
    \includegraphics[width=\textwidth]{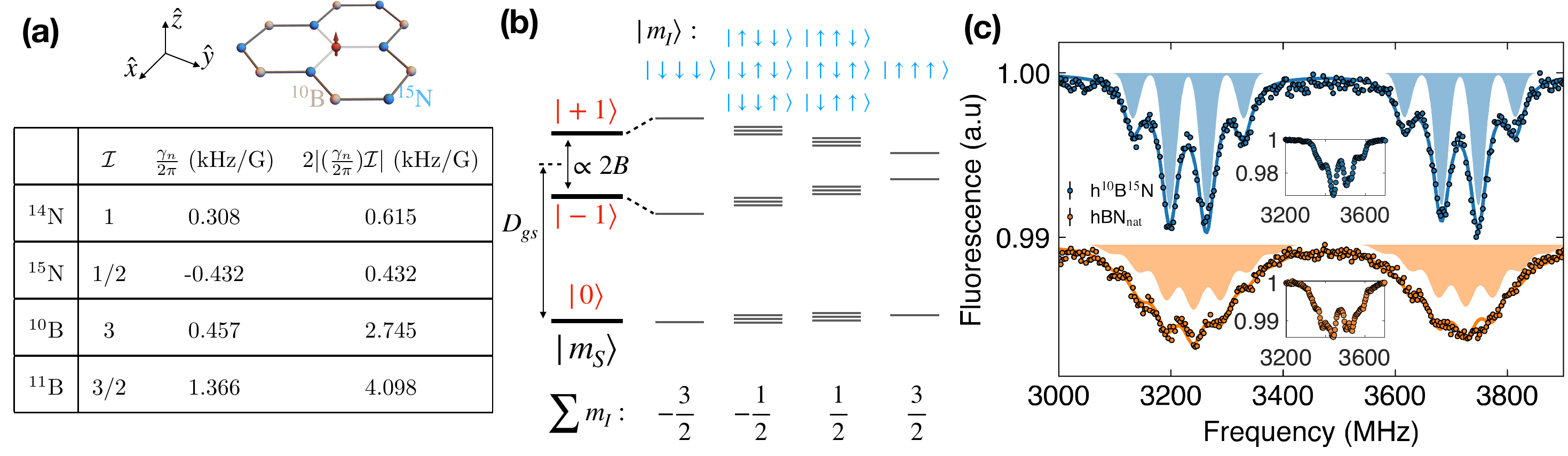}
    \caption{{\bf \vbm electron spin resonance spectra in isotopically distinct hBN samples}
    (a) Schematic of an individual \vbm center (red spin) in an isotopically purified \hbn crystal. $\hat{z}$ is defined along the c-axis (perpendicular to the lattice plane) and $\hat{x}$ and $\hat{y}$ lie in the lattice plane, with $\hat{x}$ oriented along one of the three \vbm Nitrogen bonds.
    The table lists the nuclear spin quantum number ($\mathcal{I}$), gyromagnetic ratio ($\gamma_{n}$), and overall electronic spin transition width ($\propto 2|\gamma_n \mathcal{I}|$) of four stable atomic isotopes in hBN.
    (b) Energy level diagram of the \vbm electronic ground state coupled to the three nearest-neighbor \nfive nuclear spins ($\mathcal{I}=\frac{1}{2}$) under an external magnetic field $B_z$. For each electronic spin transition $|m_s=0\rangle \leftrightarrow |m_s=\pm1\rangle$, the hyperfine interaction leads to four distinct resonances with total nuclear spin magnetic quantum number $\sum m_I = \{-\frac{3}{2}, -\frac{1}{2}, \frac{1}{2}, \frac{3}{2}\}$ of degeneracy $\{1,3,3,1\}$.
    (c) Measured ESR spectra of \vbm in \hbn and naturally abundant \hbnnat at magnetic field $B_z \approx 87~$G. Solid lines represent multi-peak Lorentzian fits, and shaded regions represent numerically simulated transitions considering the 36 nearest nuclear spins (18 nitrogen and 18 boron atoms; see Methods). Insets: ESR spectra at $B_z=0~$G.
    }
    
    \label{fig:fig1}
\end{figure*}
}

\newcommand{\figtwo}{
\begin{figure*}[t]
    \centering
    \includegraphics[width=1.0\textwidth]{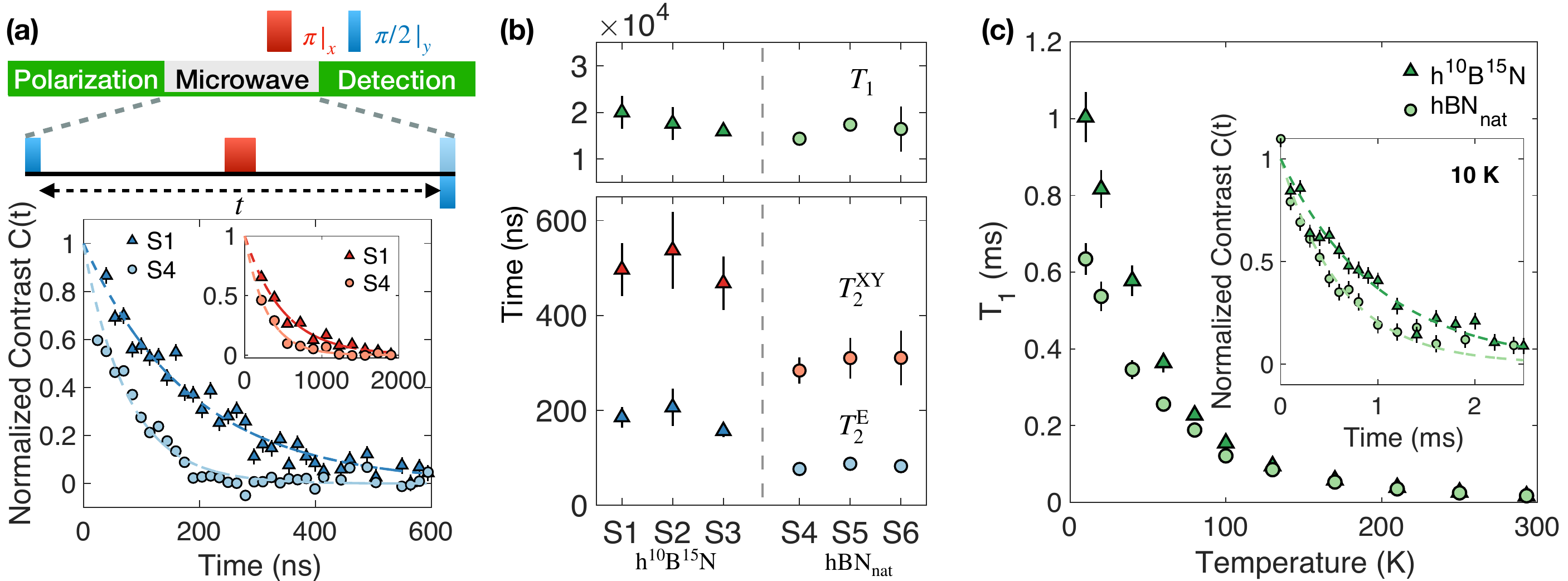}
    \caption{{\bf Coherent dynamics of \vbm in isotopically distinct hBN samples}
    (a) $T_2^\mathrm{E}$ spin echo measurements on samples S1 (\hbnns) and S4 (\hbnnatns). Insets: $T_2^\mathrm{XY}$ XY8 measurements on samples S1 and S4.
    (b) Bottom: extracted spin coherence timescales $T_2^\mathrm{E}$ and $T_2^\mathrm{XY}$. Top: extracted spin relaxation timescales $T_1$ for all six hBN samples investigated in this work.
    (c) Relaxation timescales $T_1$ for both \hbn and \hbnnat under different temperatures. Inset: $T_1$ measurement comparison with temperature at 10 K.
    }
    \label{fig:fig2}
\end{figure*}
}

\newcommand{\figthree}{
 \begin{figure}[t]
    \centering
    \includegraphics[width=\columnwidth]{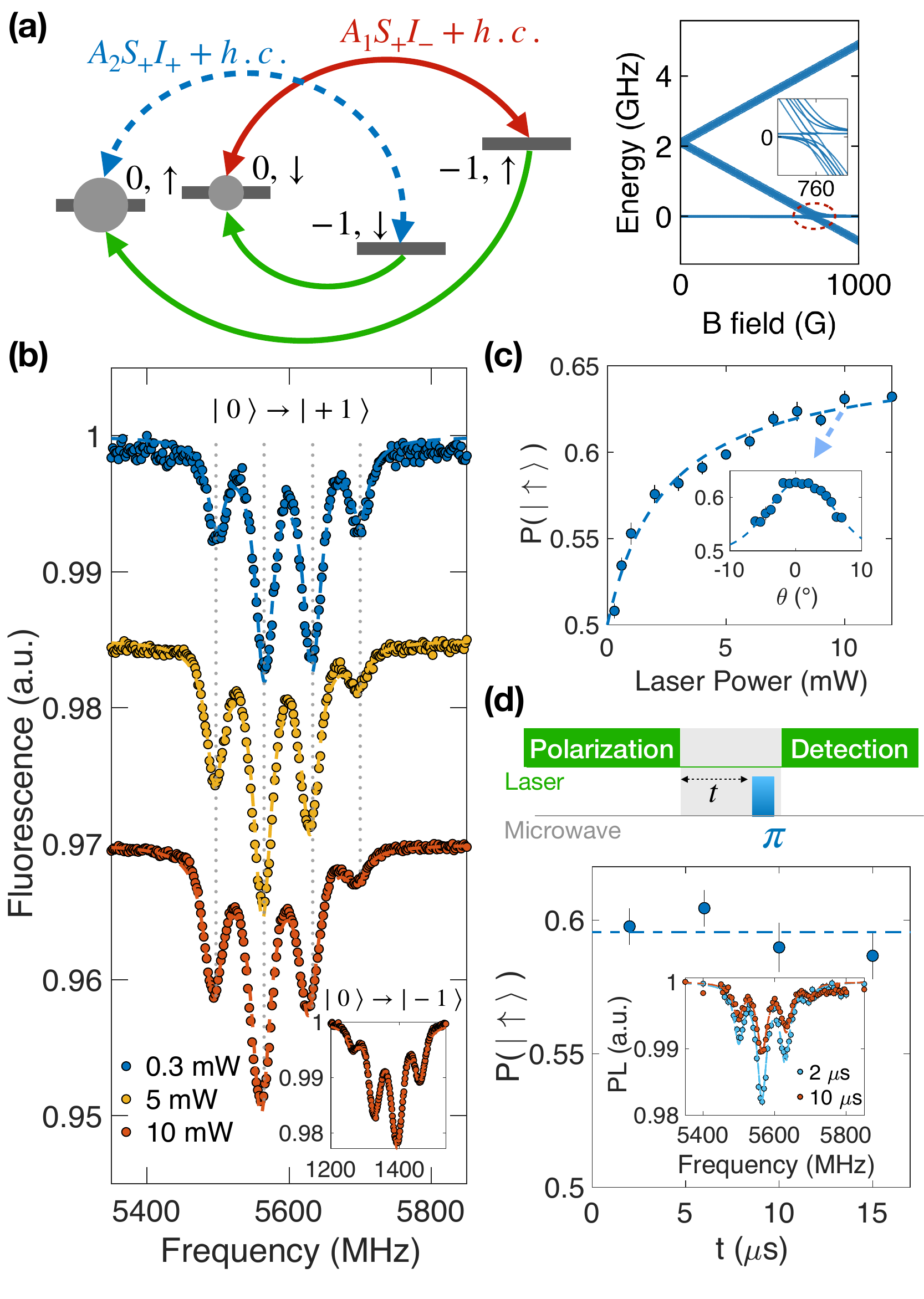}
    \caption{{\bf Dynamic polarization of nearest-neighbor \nfive nuclear spins}
    (a) Left: schematic of the nuclear spin polarization process of \nfive spins at esLAC ($B_z = 760~$G). The hyperfine term $(A_1 S_+ I_- + h.c.)$ (red) is much stronger than $(A_2 S_+ I_+ + h.c.)$ (blue). Under laser pumping (green), which continuously polarizes the electronic spin to $|m_s = 0\rangle$, each nuclear spin will be preferentially polarized to $|m_I = \:\uparrow\:\rangle$. Right: calculated excited state energy spectrum of \vbm as a function of magnetic field (see Supplementary Information). Inset: closeup of energy levels around esLAC ($760~$G).
    (b) Measured ESR spectra of the $|m_s=0\rangle \leftrightarrow|m_s=+1\rangle$ transition at esLAC under different laser powers. Inset: ESR spectrum of the $|m_s=0\rangle \leftrightarrow |m_s=-1\rangle$ transition at a laser power $\sim10~$mW. Dashed lines represent best fits assuming each \nfive nucleus is independently polarized to $|\uparrow\:\rangle$ with probability $P$.
    (c) Extracted $P(|\uparrow\:\rangle)$ as a function of laser power. Inset: $P(|\uparrow\:\rangle)$ as a function of the magnetic field alignment angle $\theta$ relative to the c-axis of hBN.
    (d) Measured depolarization dynamics of nuclear spin as a function of wait time $t$. Inset: pulsed-ESR spectra at $t = 2~\mu$s and $10~\mu$s.
    }
    \label{fig:fig3}
\end{figure}
}

\newcommand{\figfour}{
 \begin{figure}[t]
    \centering
    \includegraphics[width=1.0\columnwidth]{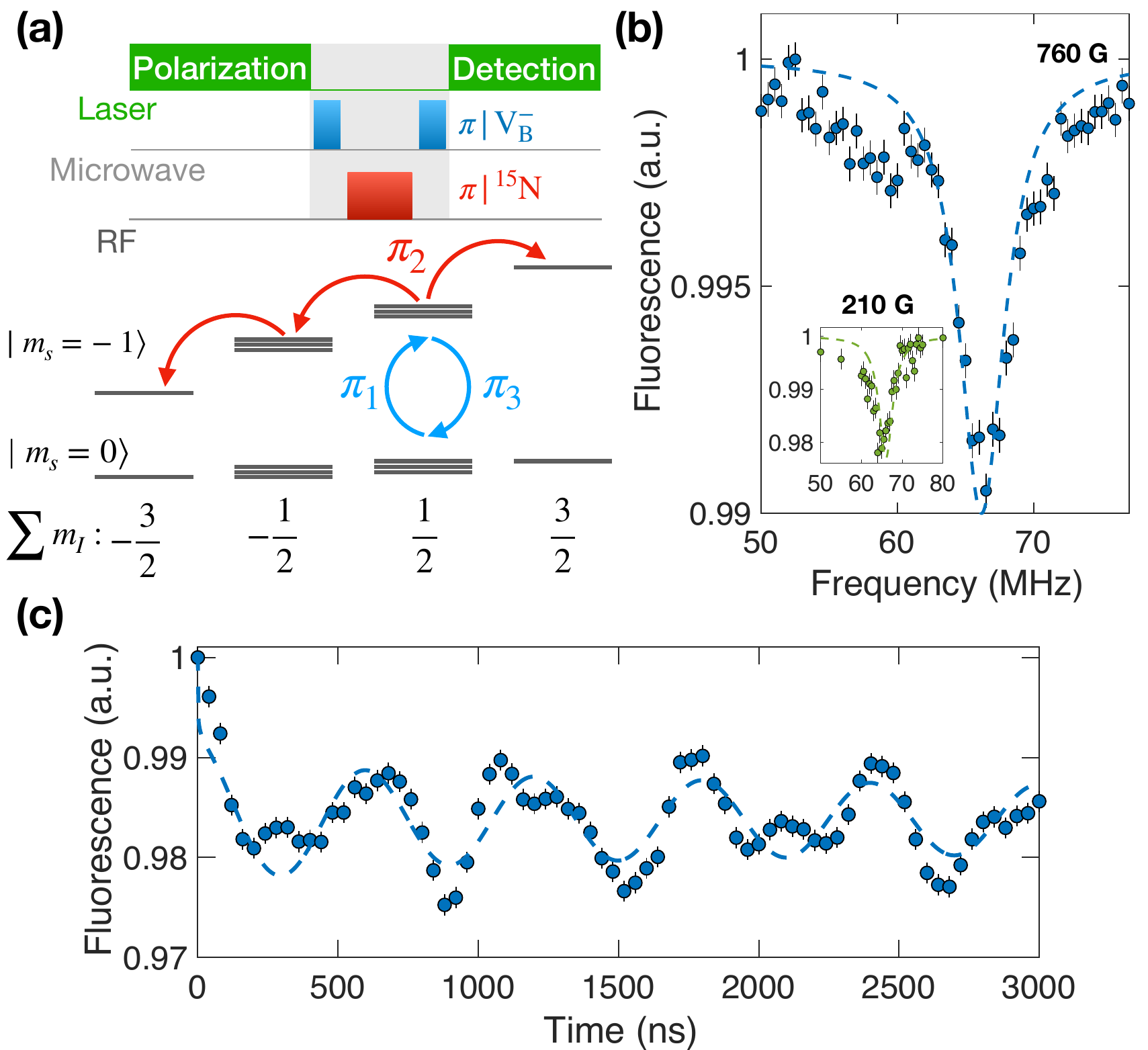}
    \caption{{\bf Coherent control of \nfive nuclear spins}
    (a) Experimental pulse sequence and schematic for the control and detection of nuclear spins.
    (b) Resulting \nfive nuclear spin resonance spectra at $760~$G and $210~$G (Inset). Dashed lines represent simulated spectra (see Supplementary Information).
    (d) Measurement of nuclear spin Rabi oscillations. The dashed line corresponds to a cosine fit with exponentially decaying amplitude.
    }
    \label{fig:fig4}
\end{figure}
}

\usepackage{graphicx}

\graphicspath{ {figures/} }

\makeatletter
\let\saved@includegraphics\includegraphics
\AtBeginDocument{\let\includegraphics\saved@includegraphics}
\makeatother

\makeatletter
\newcommand*{\centerfloat}{%
  \parindent \z@
  \leftskip \z@ \@plus 1fil \@minus \textwidth
  \rightskip\leftskip
  \parfillskip \z@skip}
\makeatother

\newcommand{\vbm}[0]{$\mathrm{V}_{\mathrm{B}}^-$ }
\newcommand{\vbmns}[0]{$\mathrm{V}_{\mathrm{B}}^-$}
\newcommand{\bten}[0]{${}^{10}\mathrm{B}$ }
\newcommand{\btenns}[0]{${}^{10}\mathrm{B}$}
\newcommand{\hbn}[0]{$\mathrm{h}{}^{10}\mathrm{B}{}^{15}\mathrm{N}$ }
\newcommand{\hbnnat}[0]{$\mathrm{hBN_\mathrm{nat}}$ }
\newcommand{\hbnns}[0]{$\mathrm{h}{}^{10}\mathrm{B}{}^{15}\mathrm{N}$}
\newcommand{\hbnnatns}[0]{$\mathrm{hBN_\mathrm{nat}}$}
\newcommand{\belev}[0]{${}^{11}\mathrm{B}$ }
\newcommand{\nfour}[0]{${}^{14}\mathrm{N}$ }
\newcommand{\nfive}[0]{${}^{15}\mathrm{N}$ }

\newcommand{\hbnelev}[0]{$\mathrm{h}{}^{11}\mathrm{B}{}^{15}\mathrm{N}$ }

\usepackage{lineno} 
\begin{document}

\title{Isotope Engineering for Spin Defects in van der Waals Materials}

\author{
Ruotian~Gong,$^{1,*}$ 
Xinyi~Du,$^{1,*}$
Eli~Janzen, $^{2}$
Vincent~Liu,$^{3}$
Zhongyuan~Liu,$^{1}$
Guanghui~He,$^{1}$
Bingtian~Ye,$^{3}$ \\
Tongcang~Li,$^{4,5}$
Norman~Y.~Yao,$^{3}$
James~H.~Edgar,$^{2}$
Erik~A.~Henriksen,$^{1,6}$ 
Chong~Zu$^{1,6,\dag}$
\\
\medskip
\normalsize{$^{1}$Department of Physics, Washington University, St. Louis, MO 63130, USA}\\
\normalsize{$^{2}$Tim Taylor Department of Chemical Engineering,
Kansas State University, Manhattan, KS 66506, USA}\\
\normalsize{$^{3}$Department of Physics, Harvard University, Cambridge, MA 02138, USA}\\
\normalsize{$^{4}$Department of Physics and Astronomy, Purdue University, West Lafayette, IN 47907, USA}\\
\normalsize{$^{5}$Elmore Family School of Electrical and Computer Engineering, Purdue University, West Lafayette, IN 47907, USA}\\
\normalsize{$^{6}$Institute of Materials Science and Engineering, Washington University, St. Louis, MO 63130, USA}\\
\normalsize{$^*$These authors contributed equally to this work}\\
\normalsize{$^\dag$To whom correspondence should be addressed; E-mail: zu@wustl.edu}\\
}

\begin{abstract}
Spin defects in van der Waals materials offer a promising platform for advancing quantum technologies. Here, we propose and demonstrate a powerful technique based on isotope engineering of host materials to significantly enhance the coherence properties of embedded spin defects. Focusing on the recently-discovered negatively charged boron vacancy center (\vbm) in hexagonal boron nitride (hBN), we grow isotopically purified \hbn crystals.
Compared to \vbm in hBN with the natural distribution of isotopes, we observe substantially narrower and less crowded \vbm spin transitions as well as extended coherence time $T_2$ and relaxation time $T_1$. 
For quantum sensing, \vbm centers in our \hbn samples exhibit a factor of $4$ ($2$) enhancement in DC (AC) magnetic field sensitivity.
For additional quantum resources, the individual addressability of the \vbm hyperfine levels enables the dynamical polarization and coherent control of the three nearest-neighbor \nfive nuclear spins.
Our results demonstrate the power of isotope engineering for enhancing the properties of quantum spin defects in hBN, and can be readily extended to improving spin qubits in a broad family of van der Waals materials.
\end{abstract}


\date{\today}

\maketitle

\emph{Introduction}--- Optically-addressable spin defects in solid-state materials have emerged as one of the leading prospects for expanding the boundary of modern quantum technologies \cite{doherty2013nitrogen,aharonovich2016solid, awschalom2018quantum, wolfowicz2021quantum, togan2010quantum,pompili2021realization,degen2017quantum,zu2021emergent,atature2018material,koehl2011room,nagy2019high, hensen2015loophole, randall2021many, hsieh2019imaging, thiel2019probing,davis2023probing, he2023quasi}.
Recently, spin defects in atomically-thin van der Waals materials have attracted significant research interest for their innate ability to integrate with heterogeneous optoelectronic and nanophotonic devices \cite{gottscholl2020initialization, gottscholl2021room,zhong2020layer, healey2023quantum,broadway2020imaging, vaidya2023quantum, gong2023coherent}.
Among a wide range of two-dimensional host materials, hexagonal boron nitride (hBN) stands out as a promising candidate owing to its large bandgap ($\sim 6$~eV) and exceptional mechanical, thermal, and chemical stability \cite{caldwell2019photonics, naclerio2023review}.
Unlike established host materials such as diamond and silicon carbide, neither nitrogen nor boron has stable isotopes with zero nuclear spin, potentially to the detriment of spin defects in hBN.
In particular, the presence of such a dense nuclear spin bath can substantially broaden electronic transitions and shorten the coherence time of spin defects.
To this end, prior experimental works have focused on designing dynamical decoupling sequences to isolate spin defects from the local nuclear spin environment in hBN \cite{gong2023coherent, rizzato2023extending}.

In this work, we demonstrate isotope engineering of hBN as a more fundamental way of strengthening the capability of spin defects for quantum applications.
By carefully optimizing the isotope species and preparing hBN samples with \nfive and \bten, we drastically improve the spin coherent properties of the recently discovered negatively charged boron-vacancy defect (\vbmns) \cite{gottscholl2020initialization, gottscholl2021room}.
%
We note that a prior study has investigated the effect of changing only boron isotopes in hBN and do not observe significant improvements in the spin properties of \vbm \cite{haykal2022decoherence}.
Importantly, this technique of isotope engineering is fundamental because of its full compatibility with other ongoing \vbm optimization efforts, making it an necessity for future \vbm improvement and applications.
Although here we focus on a specific spin defect in hBN, we would also like to highlight that the approach is exemplary for engineering other spin defects in general host materials.

\figone

\emph{Optimal Isotopes of hBN}--- 
We start with identifying the optimal isotope species in hBN for the embedded \vbm defects.
The \vbm center harbors an electronic spin triplet ground state, $|m_s = 0, \pm1\rangle$, which can be optically initialized and read out at room temperature \cite{gottscholl2020initialization}.
The Hamiltonian of the system, including both the electronic degree of freedom and nearby nuclear spins, can be written as
\begin{equation} \label{eq1}
\begin{split}
    H_\mathrm{gs} &= D_\mathrm{gs}S_z^2 +\gamma_e B_z S_z -\sum_{j}\gamma_n^j B_z I_{z}^j + \sum_{j}\mathbf{S}\mathbf{A}^j\mathbf{I}^j,
\end{split}
\end{equation}
where $D_\mathrm{gs} = (2\pi)\times 3.48~$GHz is the ground state zero-field splitting between $|m_s=0\rangle$ and $|m_s = \pm1\rangle$, $B_z$ is the external magnetic field aligned along the out-of-plane c axis of hBN ($\hat{z}$ direction), 
$\textbf{S}$ and $S_z$ are the electronic spin-1 operators, $\textbf{I}^j$ and $I_z^j$ are the spin operators for the $j^\mathrm{th}$ nuclear spin (including both nitrogen and boron) with hyperfine tensors $\mathbf{A}^j$, and $\gamma_{e} = (2\pi)\times 2.8~$MHz/G and $\gamma_n^j$ are the electronic and nuclear spin gyromagnetic ratios (Fig.~\ref{fig:fig1}a,c).
We note that, for nuclei with $\mathcal{I}\geq1$, there is a small additional nuclear spin dependent energy shift corresponding to higher multipole moments.
Nevertheless, this term only acts on the nuclear spin degree of freedom and thus has no effect on the measured electronic spin transitions to leading order. 

Throughout our experiment, we consistently operate within magnetic fields of $B_z\lesssim 760~$G such that the energy splitting $(D_\mathrm{gs} \pm \gamma_e B_z) \gtrsim (2\pi)\times 1.35~$GHz between the electronic ground state levels $|m_s = 0\rangle$ and $|m_s = \pm1\rangle$ is much greater than the hyperfine interaction strength, i.e. $|\mathbf{A}^j|\lesssim (2\pi)\times100~$MHz. 
As a result, the hyperfine term can be approximated as $\sum_{j}\mathbf{S}\mathbf{A}^j\mathbf{I}^j \approx \sum_{j}A_{zz}^j S_z I_z^j = (\sum_{j} A_{zz}^j I_z^j) S_z$ to leading order, where all remaining terms are suppressed under the secular approximation.
This term leads to an energy shift of the electronic spin transition that is dependent on the nuclear spin configuration; averaging across all different configurations gives rise to spectral crowding \cite{bourassa2020entanglement, waldherr2014quantum} of the electronic spin transition with an overall width proportional to $2|\gamma_n\mathcal{I}|$ (see Methods).

Fig.~\ref{fig:fig1}a summarizes the nuclear spin quantum numbers and gyromagnetic ratios of four stable atomic isotopes in hBN. 
Intuitively, a lower nuclear spin quantum number leads to fewer hyperfine states and contributes to less spectral crowding, with all other factors being equal.
Accordingly, the nitrogen isotope \nfive ($\mathcal{I} = 1/2$) induces a narrower transition linewidth of \vbm than its naturally abundant counterpart, \nfour ($\mathcal{I} = 1$).
However, we note that the gyromagnetic ratio is equally important in isotope engineering, and although \belev ($\mathcal{I} = 3/2$) has a lower spin quantum number compared to \bten ($\mathcal{I} = 3$), its gyromagnetic ratio is much larger, which results in a broader linewidth.
Therefore, to minimize the spectral crowding of \vbmns, we predict that \hbn is the optimal host material.

\figtwo

\emph{Experimental Characterization}--- To experimentally validate the effect of isotope engineering on the \vbm center, we grow single \hbn crystals with isotopically purified \nfive and \bten sources (purity $>99.7$~\% for \nfive and $>99.2~\%$ for \btenns, see Methods) \cite{liu2017large, liu2018single, li2020single, janzen2023boron}.
We exfoliate the hBN crystal into thin flakes with thickness ranging from $20-70~$nm.
\vbm defects are created via $\mathrm{He}^+$ ion implantation with energy 3~keV and dosage $1~\mathrm{ion}/\mathrm{nm}^{2}$, resulting an estimated \vbm concentration around $150~$ppm \cite{gong2023coherent}.
In the experiment, the fluorescence signal of \vbm is collected using a home-built confocal microscope, and the microwave is delivered via a coplanar waveguide (see Supplementary Note 1).
Most of the experiment in this work are performed at room temperature other than that specified (Fig.~\ref{fig:fig2}c).

The spin transitions of \vbm can be probed via electron spin resonance (ESR) spectroscopy: by sweeping the frequency of the applied microwave drive while monitoring the fluorescence signal of \vbm, we expect a fluorescence drop when the microwave is resonant with an electronic spin transition.
Figure~\ref{fig:fig1}c compares the ESR spectra of \vbm in our isotopically engineered \hbn and conventional naturally abundant \hbnnat ($99.6\%/0.4\%$ for ${}^{14}\mathrm{N}/{}^{15}\mathrm{N}$ and $20\%/80\%$ for ${}^{10}\mathrm{B}/{}^{11}\mathrm{B}$) under a small external magnetic field ($B\approx87~$G). 
As expected, the ESR measurements for both $|m_s=0\rangle \leftrightarrow |m_s=\pm1\rangle$ transitions on \hbn show much less crowded spectra than $\mathrm{hBN_\mathrm{nat}}$.
In particular, there are two striking features.
First, instead of an ordinary 7-resonance spectrum observed in the naturally abundant sample \cite{gottscholl2020initialization, gao2022nuclear}, we resolve 4 distinct hyperfine lines for \vbm in \hbnns.
This structure stems from the hyperfine interaction between \vbm and the three nearest-neighbor ${}^{15}\mathrm{N}$ nuclear spins $(\sum_{j=1}^{3} A_{zz}^{{}^{15}\mathrm{N}} I_z^j) S_z$, where $I_z^j$ can take the values of $\pm\frac{1}{2}$.
Accounting for all nuclear spin configurations, there are a total of 4 hyperfine lines (split by $|A_{zz}^{{}^{15}\mathrm{N}}|$) with degeneracy \{1:3:3:1\} corresponding to the total nuclear spin number $\sum m_\mathrm{I}=\{-\frac{3}{2}, -\frac{1}{2}, \frac{1}{2}, \frac{3}{2}\}$ (Fig.~\ref{fig:fig1}b,c).
By fitting the spectrum to the sum of four equally-spaced Lorentzians with amplitudes proportional to the aforementioned degeneracy ratio, we extract the hyperfine coupling $A_{zz}^{{}^{15}\mathrm{N}} = (2\pi)\times [-65.9\pm0.9]~$MHz, where the negative sign originates from the negative gyromagnetic ratio of ${}^{15}\mathrm{N}$.
In comparison, the nearest-neighbor \nfour hyperfine interaction strength in \hbnnat is $A_{zz}^{{}^{14}\mathrm{N}} = (2\pi)\times [48.3\pm0.5]~$MHz. 
The measured ratio $A_{zz}^{{}^{15}\mathrm{N}}/A_{zz}^{{}^{14}\mathrm{N}} \approx -1.36$ agrees well with the nuclear spin gyromagnetic ratios $\gamma_n^{{}^{15}\mathrm{N}}/\gamma_n^{{}^{14}\mathrm{N}}\approx -1.4$.

The second feature of the \vbm spectrum in \hbn is its dramatically narrower transitions. Specifically, the full width at half maximum (FWHM) linewidth for each hyperfine resonance is $(2\pi)\times[55\pm2]$~MHz, almost a factor of two lower than that of the naturally abundant \hbnnat sample, $(2\pi)\times[95\pm2]$~MHz.
The linewidth originates from the hyperfine couplings to boron and nitrogen nuclear spins beyond the three nearest-neighbor atoms.
To quantitatively capture the measured ESR linewidth, we perform numerical simulations by summing over the couplings to the nearest 36 nuclear spins for four different kinds of isotopically distinct hBN samples (see Methods and Supplementary Information) \cite{ivady2020ab}.
The simulated spectra for \hbn and \hbnnat are in good agreement with the experimental data (Fig.~\ref{fig:fig1}c).
Importantly, the narrower and less crowded hyperfine lines in \hbn substantially boosts the sensitivity of \vbm for quantum sensing applications \cite{gottscholl2021spin}.

\emph{Enhanced Quantum Sensing}--- Taking magnetic field sensing as an example, we now evaluate the DC and AC magnetic field sensitivity enhancement of \vbm in \hbnns.
The static magnetic field sensitivity of ESR measurement takes the form \cite{dreau2011avoiding,barry2020sensitivity}
\begin{equation} \label{eq2}
\begin{split}
\eta_\mathrm{DC} \approx\frac{2\pi}{\gamma_e \sqrt{R}} (\max|\frac{\partial{C(\nu)}}{\partial{\nu}}|)^{-1} \approx \frac{8\pi}{3\sqrt{3}} \frac{1}{\gamma_e}\frac{\Delta \nu} {C_m\sqrt{R}},
\end{split}
\end{equation} 
where $R$ denotes the photon detection rate, $C(\nu)$ the ESR measurement contrast at microwave frequency $\nu$, $C_m$ the maximum contrast, and $\Delta \nu$ the FWHM linewidth assuming a single Lorenztian resonance. 
If one directly compares the fitted FWHM from \hbn and \hbnnatns, we find a factor of $\sim 1.8$ improvement in sensitivity.
Moreover, the hyperfine transitions of \vbm in \hbnnat significantly overlap with each other, while for \vbm in \hbn the resonances are individually resolvable.
Therefore, a more accurate comparison instead takes the steepest slope, $\max|\frac{\partial{C(\nu)}}{\partial{\nu}}|$, into consideration, with which we conclude a factor of $\sim 4$ improvement in sensitivity using isotope purified \hbnns.
By optimizing the laser and microwave power, we estimate a static magnetic field sensitivity $\eta_\mathrm{DC} \approx 10~\mu$T~$\mathrm{Hz}^{-\frac{1}{2}}$. We note that the sensitivities reported in this work are sample specific, and one can further optimize the sensitivity by increasing the number of \vbm defects used in experiment (see Supplementary Information).

Next, we characterize the spin echo coherence of \vbm in \hbn and \hbnnat for AC field sensing.
The spin echo coherent timescale, $T_2^\mathrm{E}$, of \vbm has been previously understood to be limited by the magnetic fluctuations from the nearby nuclear spin bath \cite{gong2023coherent, yang2008quantum, haykal2022decoherence}, and the isotope engineering technique should precisely alleviate such decoherence.
For a robust and systematic comparison, we prepare three \hbn flakes (sample S1-S3) and three naturally abundant \hbnnat flakes (sample S4-S6) onto the same microwave stripline (see Methods).
Figure~\ref{fig:fig2}a shows the measured spin echo decays on \hbn sample S1 and \hbnnat sample S4, where the extension of spin echo timescales $T_2^\mathrm{E}$ exceeds twofold, from $T_\mathrm{2,S4}^\mathrm{E}=87\pm9~$ns to $T_\mathrm{2,S1}^\mathrm{E}=186\pm22~$ns.
Figure~\ref{fig:fig2}b summarizes the experimentally measured spin echo coherence timescales $T_2^\mathrm{E}$ across all six hBN flakes, revealing a consistent improvement of coherence time in our isotopically purified samples.
We also perform the spin echo measurement using \hbnelev samples and find $T_2^\mathrm{E}\approx 119\pm6~$ns, lying in between \hbn and \hbnnat (see Supplementary Fig.~13). 
\figthree

The AC magnetic field sensitivity $\eta_\mathrm{AC}\propto T_2^{-1}$ exhibits improvement of more than a factor of $2$ in our \hbn samples.
By employing an advanced dynamical decoupling sequence, XY8, which exploits a series of echo pulses with alternating phases, we further extend the coherence time of \vbm in \hbn samples to $T_2^\mathrm{XY} \approx 500$~ns (Fig.~\ref{fig:fig2}b Inset).
This improves the estimated AC magnetic field sensitivity to $\eta_\mathrm{AC} \approx 7~\mu$T~$\mathrm{Hz}^{-\frac{1}{2}}$ for a signal at frequency $\sim (2\pi)\times12~$MHz (see Supplementary Note 4). 

We now turn to investigate the spin relaxation time, $T_1$, of \vbm in different isotopic samples.
At room temperature, we find $T_1$ of \vbm across all six samples to be comparable ($T_1\approx 15~\mu$s) and nearly independent of isotope choice (Fig.~\ref{fig:fig2}b).
To further explore the isotope effect on $T_1$, we also explore the temperature-dependence of the \vbm relaxation using an optical cryostat~(Fig.~\ref{fig:fig2}c).
The measured $T_1$ in both \hbn and \hbnnat increases monotonically with decreasing temperatures.
Interestingly, when the temperature goes below $\sim170$~K, there is a clear improvement of $T_1$ in the \hbn sample compared to natural abundant sample.
At the lowest temperature $10~$K, we find $T_1 = (1.00 \pm 0.07)$~ms for \vbm in \hbnns, around $50~\%$ longer than $T_1 = (0.63\pm 0.04)$~ms measured in \hbnnat
(Fig.~\ref{fig:fig2}c Inset).
The current theoretical understanding of the spin relaxation process of \vbm identifies spin-phonon interaction as the main limitation \cite{gottscholl2021room, durand2023optically, cambria2023temperature}, which is consistent with the observed monotonic increase of $T_1$ with decreasing temperature in both samples. 
One possible explanation for the improved $T_1$ of \vbm in \hbn is that the nearest-neighbouring \nfive nuclei are slightly heavier in mass, leading to a weaker spin-phonon coupling strength \cite{tang2023first, mondal2023spin}.
However, the detailed underlying mechanism of isotope and temperature dependent of \vbm spin lifetime invites more future studies.
Nevertheless, the extension of $T_1$ from isotope effect facilitates the use of \vbm as a noise magnetometer to diagnose different phases of material, such as magnetic insulators and superconductivity at low temperature\cite{hsieh2019imaging,chatterjee2019diagnosing, chatterjee2022single, huang2022wide}.

\emph{Polarizing Three Nearest \nfive Nuclear Spins}--- 
Next, we demonstrate the polarization and coherent control of the three nearest-neighbor \nfive nuclear spins using \vbmns. 
Nuclear spins feature exceptional isolation from external environments and offer long-lived systems for quantum simulation and computation applications.
In contrast to \vbm in naturally abundant hBN samples, our isotopically purified host \hbn presents a unique advantage: the hyperfine levels are individually addressable with a much less crowded spectrum.

We start by dynamically polarizing the \nfive nuclear spins at the electronic spin level anti-crossing (esLAC).
Under an external magnetic field $B_z \approx 760~$G, the \vbm excited state levels $|m_s = 0\rangle$ and $|m_s = -1\rangle$ are nearly degenerate \cite{mathur2022excited, baber2021excited, yu2022excited}.
In this regime, the secular approximation no longer holds, and we need to consider the full hyperfine interaction Hamiltonian,
\begin{equation} \label{eq3}
\begin{split}
    \sum_{j=1}^{3}\mathbf{S}\mathbf{A}^{j}\mathbf{I}^j & = \sum_{j=1}^{3} (A_{zz}^{j}S_z I_z^j + A_{xx}^{j}S_x I_x^j  \\
& +A_{yy}^{j}S_y I_y^j +A_{xy}^{j}S_xI_y^j +A_{yx}^{j}S_yI_x^j ).
\end{split}
\end{equation}
To understand the nuclear spin polarization process, we can rewrite the hyperfine terms using ladder operators  as
\begin{equation} \label{eq4}
\begin{split}
    \sum_{j=1}^{3}\mathbf{S}\mathbf{A}^{j}\mathbf{I}^j & = \sum_{j=1}^{3} [A_{zz}^{j}S_z I_z^j + (A_1^{j}S_+ I_-^j + h.c.) \\ & + (A_2^{j}S_+ I_+^j + h.c.)],
\end{split}
\end{equation}
where $A_1^j = \frac{1}{4}(A_{xx}^j+A_{yy}^j)$, $A_2^j = \frac{1}{4} (A_{xx}^j-A_{yy}^j)+\frac{1}{2i}A_{xy}^j$, $S_{\pm}$ and $I_{\pm}$ are the ladder operators (see Supplementary Information).

From Eq.~\ref{eq4}, the nuclear spin polarization process at esLAC is made apparent. 
Specifically, the term $(A_1^{j}S_+ I_-^j + h.c.)$ leads to electron-nuclear spin ``flip-flop", $|m_s=0, m_I=
\:\downarrow \rangle \leftrightarrow |m_s=-1, m_I=\:\uparrow\rangle$, while the term $(A_2^{j}S_+ I_+^j + h.c.)$ connects the other two states, $|m_s=0, m_I=\:\uparrow \rangle \leftrightarrow |m_s=-1, m_I=\:\downarrow \rangle$ (Fig.~\ref{fig:fig3}a).
From previous ab-initio calculations \cite{ivady2020ab,gao2022nuclear}, $|A_1^{j}| > |A_2^{j}|$ for the three nearest-neighbor \nfive nuclear spins. 
Therefore, under the strong optical polarization that continuously pumps the electronic spin from $|m_s=\pm1\rangle$ to $|m_s=0\rangle$ (while leaving the nuclear spin unchanged), each \nfive will be preferentially polarized to $|m_I = \:\uparrow\rangle$. 

To experimentally probe the nuclear spin polarization, we measure ESR spectra at a range of different laser powers.
Here, we focus on the ESR transitions from $|m_s = 0\rangle$ to $|m_s = +1\rangle$ to avoid any fluorescence modulation due to esLAC and thus accurately characterize the nuclear spin population \cite{zu2014experimental}.
At low laser power ($0.3~$mW), the ESR spectrum exhibits a symmetric four-peak structure with amplitudes \{1:3:3:1\}, similar to the low field measurement, indicating no nuclear spin polarization (Fig.~\ref{fig:fig3}b).
Once the laser power, and hence the optical pumping rate, increases, the ESR spectrum manifests an asymmetry skewed toward the two resonances at lower frequencies.
This indicates that the three nearest-neighbor \nfive nuclear spins are significantly polarized to the $|m_I =\: \uparrow\rangle$ state.

\figfour

To quantify the nuclear spin polarization, we assume each nuclear spin is independently initialized to $|m_I=\:\uparrow \rangle$ with probability $P(\ket{\uparrow})$ and $|m_I = \:\downarrow\rangle$ with probability $1-P$.
In this case, the population distribution of the four hyperfine resonances follows a ratio of $\{P^3 :3 P^2(1-P) : 3 P(1-P)^2 : (1-P)^3\}$ (See Methods).
This model yields excellent agreement with the experimental data (Fig.~\ref{fig:fig3}b), allowing us to extract the nuclear spin polarization probability $P$ at different laser powers.
The nuclear polarization saturates at a laser power of around $10~$mW, giving a maximum extracted nuclear spin polarization probability $P = (63.2\pm0.3)\%$ (Fig.~\ref{fig:fig3}c).
The nuclear spin polarization is also sensitive to the magnetic field alignment relative to the c-axis of the \hbnns, as a transverse magnetic field induces mixing between the nuclear spin states (Fig.~\ref{fig:fig3}c inset).
We notice a small overall shift ($\sim 5~$MHz) of the entire ESR spectrum at high laser power originating from the laser-induced heating effect of the sample \cite{gottscholl2021spin}. 
Nevertheless, the well-resolved hyperfine lines in our isotopically purified \hbn enables us to faithfully capture any temperature-induced shift that may have been previously difficult to distinguish in the spectrum of naturally abundant \hbnnatns.

We further study the nuclear spin depolarization dynamics via pulsed ESR measurements.
After initializing the \vbm electronic spin and the three \nfive nuclear spins at esLAC with a $5~\mu$s laser pulse, we wait for a time $t$ before applying a microwave $\pi$-pulse with varying frequency to probe the hyperfine transitions, as shown in Figure~\ref{fig:fig3}d.
By fitting the ESR spectra, we observe that the nuclear spin population $P$ exhibits almost no decay within 15 $\mu$s, the $T_1$ time of $V_\mathrm{B}^{-}$. 
The longevity of nuclear spin polarization may serve as a vital attribute for future applications such as quantum registers \cite{bradley2019ten}.

\emph{Coherent Control of the Nuclear Spins}--- Next, we illustrate the direct control and detection of the three nearest-neighbor \nfive nuclear spins.
After polarizing the \vbm to $|m_s=0\rangle$ state via optical pumping, we apply a nuclear spin selective microwave $\pi$-pulse to transfer $|m_s = 0, \sum m_I = \frac{1}{2}\rangle$ population into $|m_s = -1, \sum m_I = \frac{1}{2}\rangle$ followed by a radio-frequency (RF) $\pi$-pulse to drive the nuclear spin states (Fig.~\ref{fig:fig4}a).
A final microwave $\pi$-pulse transfers $|m_s = -1, \sum m_I = \frac{1}{2}\rangle$ back to $|m_s = 0, \sum m_I = \frac{1}{2}\rangle$ for nuclear spin detection.

Figure~\ref{fig:fig4}b shows the measured \nfive nuclear spin resonance spectrum obtained by sweeping the frequency of the RF pulse.
The clear nuclear spin transition at around $(2\pi)\times66.6~$MHz is consistent with the sum of the measured \nfive hyperfine interaction strength and the nuclear spin Zeeman shift $|A^{{}^{15}\mathrm{N}}_{zz}+\gamma_n B_z| \approx (2\pi)\times66.2~$MHz.
By parking the RF pulse at this frequency and increasing the pulse duration, we obtain coherent Rabi oscillations of the nearest three \nfive nuclear spins with Rabi frequency $\Omega_\mathrm{N} = (2\pi)\times (1.67\pm0.02)~$MHz (Fig.~\ref{fig:fig4}c).

Here, we would like to emphasize a few points. 
First, the nuclear spin Rabi signal contains a clear beating that does not fit to a single frequency oscillation.
This originates from the interaction between nuclear spins which can be captured using a more sophisticated model containing \vbm and three \nfive nuclear spins (see Supplementary Note 6.3).
Second, the transverse hyperfine interaction enhances the effective nuclear gyromagnetic ratio to \cite{sangtawesin2016hyperfine, gao2022nuclear}
\begin{equation}
    \gamma_n^\mathrm{eff} \approx \dfrac{\gamma_e \sqrt{A^2_{xx}+A^2_{yy}+2A^2_{xy}}}{\sqrt{2}(D_\mathrm{gs}-\gamma_e B_z)},
\end{equation}
enabling us to achieve a fast nuclear spin manipulation (see Supplementary Note 6.1).
From our experiment, we estimate $\gamma_n^\mathrm{eff}/\gamma_n\approx 99$, suggesting $\sqrt{A^2_{xx}+A^2_{yy}+2A^2_{xy}} \approx (2\pi)\times 30~$MHz, a factor approximately 5 times lower than the values predicted by ab-initio calculations \cite{ivady2020ab, gao2022nuclear} (see Supplementary Note 6.2).
Another piece of evidence is that the measured nuclear spin resonance spectrum is consistent with the estimated transverse hyperfine parameters (Fig.~\ref{fig:fig4}b), while if we include the ab-initio predicted values, the simulated spectrum deviates from the measurement at $B\approx 760~$G (Supplementary Fig.~9).
Prior work using \nfour nuclear spins in natural abundant \hbnnat has reported a $\sim30~\%$ discrepancy between experiment and the ab-initio calculated transverse hyperfine coefficients~\cite{gao2022nuclear}.
However, the measured \nfour nuclear spin transitions contain multiple resonances and are much broader than what we have observed here using \nfive nuclei.
Future work is required to resolve such discrepancy between experiment and ab-initio results.
For instance, one can perform high-resolution ESR spectroscopy near the ground-state anti-crossing (gsLAC) with magnetic field $B\approx 1200~$G. 
At such field, the \vbm electronic spin state $|0\rangle$ and $|-1\rangle$ are close to degenerate, and one expects the effect of transverse hyperfine coefficients to become evident.

\emph{Outlook}--- In conclusion, we present isotope engineering as a versatile and foundational tool to improving spin properties of \vbmns.
On the quantum sensing front, \vbm centers in isotope-engineered \hbn feature substantially narrower spin transitions and enhanced spin coherent time
compared to conventional \hbnnatns, enabling a universal enhancement factor of $\sim 4(2)$ in DC (AC) magnetic field sensitivity.
Additionally, we demonstrate the dynamical polarization and coherent control of the nearest neighboring three \nfive nuclear spins.

Looking forward, our work also opens the door to a few highly promising directions. 
First, we believe that \emph{almost all} previous works using \vbm for sensing applications could greatly benefit from simply switching to our newly prepared \hbn sample.
The demonstrated sensitivity improvement is also fully compatible with other ongoing optimization efforts on \vbm sensors, such as dynamical decoupling protocols to extend $T_2$ \cite{gong2023coherent,rizzato2023extending}, coupling \vbm to optical cavities \cite{fröch2021coupling, mendelson2022coupling, qian2022unveiling} and increasing \vbm number and density \cite{gong2023coherent,whitefield2023magnetic, zhou2023dc} to improve fluorescent signals.
Second, the individual addressability of \vbm hyperfine transitions in \hbn allows the utilization of nearby nuclear spins as valuable quantum resources and provides a promising platform to realize multi-qubit quantum registers in two-dimensional materials.
Finally, we highlight that our isotope engineering method can be readily extended to other spin defects in hBN \cite{guo2023coherent, stern2022room} as well as the broader family of van der Waals materials \cite{li2022carbon}.
For instance, the recent search for quantum spin defects in transition metal dichalcogenides \cite{lee2022spin} will greatly benefit from a careful selection of host isotopes, which may even create a nuclear spin-free environment.

\vspace{2mm}

\emph{Note added}: During the completion of this work, we became aware of complementary work studying boron-vacancy centers in both nitrogen and boron isotopes purified hBN crystals (Vincent Jacques, private communication)~\cite{jacques2023isotope}.

\vspace{1mm}

\emph{Acknowledgements}: We gratefully acknowledge Vincent~Jacques, Cassabois~Guillaume, Bernard~Gil, Cong~Su, Emily~Davis, Weijie~Wu, Wonjae~Lee, Khanh~Pham, Benchen~Huang, Xingrui~Song, Li~Yang, Du~Li, Xingyu~Gao, and Joonhee~Choi for helpful discussions. We thank Justin S. Kim, Zhihao Xu, and Sang-Hoon Bae for their assistance in experiment. 
This work is supported by the Startup Fund, the Center for Quantum Leaps, the Institute of Materials Science and Engineering, and the OVCR Seed Grant from Washington University. 
T.~Li acknowledges support from the Gordon and Betty Moore Foundation grant 10.37807/gbmf12259.
V.~Liu, B.~Ye, and N.~Y.~Yao acknowledge support from the U.S. Department of Energy through BES grant no. DE-SC0019241 and through the DOE Office of Science, Office of Basic Energy Sciences, Materials Sciences and Engineering Division and the Division of Chemical Sciences, Geosciences and Biosciences at LBNL under Contract no. DE-AC02-05-CH11231.  
Support of E.~Janzen and J.~H.~Edgar for hBN crystal growth  is provided by the Office of Naval Research, award number N00014-22-1-2582.
X.\ Du and E.\ A.\ Henriksen acknowledge support from the Moore Foundation Experimental Physics Investigators Initiative award no.\ 11560.

\vspace{1mm}

\emph{Author contributions}--- C.Z. conceived the idea. R.G., X.D., Z.L. and G.H. performed the experiment. R.G., X.D., V.L., Z.L., B.Y., N.Y. and C.Z. developed the theoretical models and performed the numerical simulations. R.G., X.D., V.L., Z.L. and C.Z. performed the data analysis. E.J. and J.H.E. grew the hBN samples. R.G., X.D., T.L. and E.A.H. fabricated the microwave delivery system. E.A.H. and C.Z. supervised the project. R.G., X.D., V.L. and C.Z. wrote the manuscript with inputs from all authors.

\bibliographystyle{naturemag}
\bibliography{ref}

\begin{thebibliography}{10}
\expandafter\ifx\csname url\endcsname\relax
  \def\url#1{\texttt{#1}}\fi
\expandafter\ifx\csname urlprefix\endcsname\relax\def\urlprefix{URL }\fi
\providecommand{\bibinfo}[2]{#2}
\providecommand{\eprint}[2][]{\url{#2}}

\bibitem{doherty2013nitrogen}
\bibinfo{author}{Doherty, M.~W.} \emph{et~al.}
\newblock \bibinfo{title}{The nitrogen-vacancy colour centre in diamond}.
\newblock \emph{\bibinfo{journal}{Physics Reports}}
  \textbf{\bibinfo{volume}{528}}, \bibinfo{pages}{1--45}
  (\bibinfo{year}{2013}).

\bibitem{aharonovich2016solid}
\bibinfo{author}{Aharonovich, I.}, \bibinfo{author}{Englund, D.} \&
  \bibinfo{author}{Toth, M.}
\newblock \bibinfo{title}{Solid-state single-photon emitters}.
\newblock \emph{\bibinfo{journal}{Nature Photonics}}
  \textbf{\bibinfo{volume}{10}}, \bibinfo{pages}{631--641}
  (\bibinfo{year}{2016}).

\bibitem{awschalom2018quantum}
\bibinfo{author}{Awschalom, D.~D.}, \bibinfo{author}{Hanson, R.},
  \bibinfo{author}{Wrachtrup, J.} \& \bibinfo{author}{Zhou, B.~B.}
\newblock \bibinfo{title}{Quantum technologies with optically interfaced
  solid-state spins}.
\newblock \emph{\bibinfo{journal}{Nature Photonics}}
  \textbf{\bibinfo{volume}{12}}, \bibinfo{pages}{516--527}
  (\bibinfo{year}{2018}).

\bibitem{wolfowicz2021quantum}
\bibinfo{author}{Wolfowicz, G.} \emph{et~al.}
\newblock \bibinfo{title}{Quantum guidelines for solid-state spin defects}.
\newblock \emph{\bibinfo{journal}{Nature Reviews Materials}}
  \textbf{\bibinfo{volume}{6}}, \bibinfo{pages}{906--925}
  (\bibinfo{year}{2021}).

\bibitem{togan2010quantum}
\bibinfo{author}{Togan, E.} \emph{et~al.}
\newblock \bibinfo{title}{Quantum entanglement between an optical photon and a
  solid-state spin qubit}.
\newblock \emph{\bibinfo{journal}{Nature}} \textbf{\bibinfo{volume}{466}},
  \bibinfo{pages}{730--734} (\bibinfo{year}{2010}).

\bibitem{pompili2021realization}
\bibinfo{author}{Pompili, M.} \emph{et~al.}
\newblock \bibinfo{title}{Realization of a multinode quantum network of remote
  solid-state qubits}.
\newblock \emph{\bibinfo{journal}{Science}} \textbf{\bibinfo{volume}{372}},
  \bibinfo{pages}{259--264} (\bibinfo{year}{2021}).

\bibitem{degen2017quantum}
\bibinfo{author}{Degen, C.~L.}, \bibinfo{author}{Reinhard, F.} \&
  \bibinfo{author}{Cappellaro, P.}
\newblock \bibinfo{title}{Quantum sensing}.
\newblock \emph{\bibinfo{journal}{Reviews of modern physics}}
  \textbf{\bibinfo{volume}{89}}, \bibinfo{pages}{035002}
  (\bibinfo{year}{2017}).

\bibitem{zu2021emergent}
\bibinfo{author}{Zu, C.} \emph{et~al.}
\newblock \bibinfo{title}{Emergent hydrodynamics in a strongly interacting
  dipolar spin ensemble}.
\newblock \emph{\bibinfo{journal}{Nature}} \textbf{\bibinfo{volume}{597}},
  \bibinfo{pages}{45--50} (\bibinfo{year}{2021}).

\bibitem{atature2018material}
\bibinfo{author}{Atat{\"u}re, M.}, \bibinfo{author}{Englund, D.},
  \bibinfo{author}{Vamivakas, N.}, \bibinfo{author}{Lee, S.-Y.} \&
  \bibinfo{author}{Wrachtrup, J.}
\newblock \bibinfo{title}{Material platforms for spin-based photonic quantum
  technologies}.
\newblock \emph{\bibinfo{journal}{Nature Reviews Materials}}
  \textbf{\bibinfo{volume}{3}}, \bibinfo{pages}{38--51} (\bibinfo{year}{2018}).

\bibitem{koehl2011room}
\bibinfo{author}{Koehl, W.~F.}, \bibinfo{author}{Buckley, B.~B.},
  \bibinfo{author}{Heremans, F.~J.}, \bibinfo{author}{Calusine, G.} \&
  \bibinfo{author}{Awschalom, D.~D.}
\newblock \bibinfo{title}{Room temperature coherent control of defect spin
  qubits in silicon carbide}.
\newblock \emph{\bibinfo{journal}{Nature}} \textbf{\bibinfo{volume}{479}},
  \bibinfo{pages}{84--87} (\bibinfo{year}{2011}).

\bibitem{nagy2019high}
\bibinfo{author}{Nagy, R.} \emph{et~al.}
\newblock \bibinfo{title}{High-fidelity spin and optical control of single
  silicon-vacancy centres in silicon carbide}.
\newblock \emph{\bibinfo{journal}{Nature Communications}}
  \textbf{\bibinfo{volume}{10}}, \bibinfo{pages}{1954} (\bibinfo{year}{2019}).

\bibitem{hensen2015loophole}
\bibinfo{author}{Hensen, B.} \emph{et~al.}
\newblock \bibinfo{title}{Loophole-free bell inequality violation using
  electron spins separated by 1.3 kilometres}.
\newblock \emph{\bibinfo{journal}{Nature}} \textbf{\bibinfo{volume}{526}},
  \bibinfo{pages}{682--686} (\bibinfo{year}{2015}).

\bibitem{randall2021many}
\bibinfo{author}{Randall, J.} \emph{et~al.}
\newblock \bibinfo{title}{Many-body--localized discrete time crystal with a
  programmable spin-based quantum simulator}.
\newblock \emph{\bibinfo{journal}{Science}} \textbf{\bibinfo{volume}{374}},
  \bibinfo{pages}{1474--1478} (\bibinfo{year}{2021}).

\bibitem{hsieh2019imaging}
\bibinfo{author}{Hsieh, S.} \emph{et~al.}
\newblock \bibinfo{title}{Imaging stress and magnetism at high pressures using
  a nanoscale quantum sensor}.
\newblock \emph{\bibinfo{journal}{Science}} \textbf{\bibinfo{volume}{366}},
  \bibinfo{pages}{1349--1354} (\bibinfo{year}{2019}).

\bibitem{thiel2019probing}
\bibinfo{author}{Thiel, L.} \emph{et~al.}
\newblock \bibinfo{title}{Probing magnetism in 2d materials at the nanoscale
  with single-spin microscopy}.
\newblock \emph{\bibinfo{journal}{Science}} \textbf{\bibinfo{volume}{364}},
  \bibinfo{pages}{973--976} (\bibinfo{year}{2019}).

\bibitem{davis2023probing}
\bibinfo{author}{Davis, E.} \emph{et~al.}
\newblock \bibinfo{title}{Probing many-body noise in a strongly interacting
  two-dimensional dipolar spin system}.
\newblock \emph{\bibinfo{journal}{Nature Physics}}  (\bibinfo{year}{2023}).

\bibitem{he2023quasi}
\bibinfo{author}{He, G.} \emph{et~al.}
\newblock \bibinfo{title}{Quasi-floquet prethermalization in a disordered
  dipolar spin ensemble in diamond}.
\newblock \emph{\bibinfo{journal}{Physical Review Letters}}
  \textbf{\bibinfo{volume}{131}}, \bibinfo{pages}{130401}
  (\bibinfo{year}{2023}).

\bibitem{gottscholl2020initialization}
\bibinfo{author}{Gottscholl, A.} \emph{et~al.}
\newblock \bibinfo{title}{Initialization and read-out of intrinsic spin defects
  in a van der waals crystal at room temperature}.
\newblock \emph{\bibinfo{journal}{Nature Materials}}
  \textbf{\bibinfo{volume}{19}}, \bibinfo{pages}{540--545}
  (\bibinfo{year}{2020}).

\bibitem{gottscholl2021room}
\bibinfo{author}{Gottscholl, A.} \emph{et~al.}
\newblock \bibinfo{title}{Room temperature coherent control of spin defects in
  hexagonal boron nitride}.
\newblock \emph{\bibinfo{journal}{Science Advances}}
  \textbf{\bibinfo{volume}{7}}, \bibinfo{pages}{eabf3630}
  (\bibinfo{year}{2021}).

\bibitem{zhong2020layer}
\bibinfo{author}{Zhong, D.} \emph{et~al.}
\newblock \bibinfo{title}{Layer-resolved magnetic proximity effect in van der
  waals heterostructures}.
\newblock \emph{\bibinfo{journal}{Nature Nanotechnology}}
  \textbf{\bibinfo{volume}{15}}, \bibinfo{pages}{187--191}
  (\bibinfo{year}{2020}).

\bibitem{healey2023quantum}
\bibinfo{author}{Healey, A.} \emph{et~al.}
\newblock \bibinfo{title}{Quantum microscopy with van der waals
  heterostructures}.
\newblock \emph{\bibinfo{journal}{Nature Physics}}
  \textbf{\bibinfo{volume}{19}}, \bibinfo{pages}{87--91}
  (\bibinfo{year}{2023}).

\bibitem{broadway2020imaging}
\bibinfo{author}{Broadway, D.~A.} \emph{et~al.}
\newblock \bibinfo{title}{Imaging domain reversal in an ultrathin van der waals
  ferromagnet}.
\newblock \emph{\bibinfo{journal}{Advanced Materials}}
  \textbf{\bibinfo{volume}{32}}, \bibinfo{pages}{2003314}
  (\bibinfo{year}{2020}).

\bibitem{vaidya2023quantum}
\bibinfo{author}{Vaidya, S.}, \bibinfo{author}{Gao, X.},
  \bibinfo{author}{Dikshit, S.}, \bibinfo{author}{Aharonovich, I.} \&
  \bibinfo{author}{Li, T.}
\newblock \bibinfo{title}{Quantum sensing and imaging with spin defects in
  hexagonal boron nitride}.
\newblock \emph{\bibinfo{journal}{Advances in Physics: X}}
  \textbf{\bibinfo{volume}{8}}, \bibinfo{pages}{2206049}
  (\bibinfo{year}{2023}).

\bibitem{gong2023coherent}
\bibinfo{author}{Gong, R.} \emph{et~al.}
\newblock \bibinfo{title}{Coherent dynamics of strongly interacting electronic
  spin defects in hexagonal boron nitride}.
\newblock \emph{\bibinfo{journal}{Nature Communications}}
  \textbf{\bibinfo{volume}{14}}, \bibinfo{pages}{3299} (\bibinfo{year}{2023}).

\bibitem{caldwell2019photonics}
\bibinfo{author}{Caldwell, J.~D.} \emph{et~al.}
\newblock \bibinfo{title}{Photonics with hexagonal boron nitride}.
\newblock \emph{\bibinfo{journal}{Nature Reviews Materials}}
  \textbf{\bibinfo{volume}{4}}, \bibinfo{pages}{552--567}
  (\bibinfo{year}{2019}).

\bibitem{naclerio2023review}
\bibinfo{author}{Naclerio, A.~E.} \& \bibinfo{author}{Kidambi, P.~R.}
\newblock \bibinfo{title}{A review of scalable hexagonal boron nitride (h-bn)
  synthesis for present and future applications}.
\newblock \emph{\bibinfo{journal}{Advanced Materials}}
  \textbf{\bibinfo{volume}{35}}, \bibinfo{pages}{2207374}
  (\bibinfo{year}{2023}).

\bibitem{rizzato2023extending}
\bibinfo{author}{Rizzato, R.} \emph{et~al.}
\newblock \bibinfo{title}{Extending the coherence of spin defects in hbn
  enables advanced qubit control and quantum sensing}.
\newblock \emph{\bibinfo{journal}{Nature Communications}}
  \textbf{\bibinfo{volume}{14}}, \bibinfo{pages}{5089} (\bibinfo{year}{2023}).

\bibitem{haykal2022decoherence}
\bibinfo{author}{Haykal, A.} \emph{et~al.}
\newblock \bibinfo{title}{Decoherence of $\mathrm{V}_\mathrm{B}^-$ spin defects
  in monoisotopic hexagonal boron nitride}.
\newblock \emph{\bibinfo{journal}{Nature Communications}}
  \textbf{\bibinfo{volume}{13}}, \bibinfo{pages}{4347} (\bibinfo{year}{2022}).

\bibitem{bourassa2020entanglement}
\bibinfo{author}{Bourassa, A.} \emph{et~al.}
\newblock \bibinfo{title}{Entanglement and control of single nuclear spins in
  isotopically engineered silicon carbide}.
\newblock \emph{\bibinfo{journal}{Nature Materials}}
  \textbf{\bibinfo{volume}{19}}, \bibinfo{pages}{1319--1325}
  (\bibinfo{year}{2020}).

\bibitem{waldherr2014quantum}
\bibinfo{author}{Waldherr, G.} \emph{et~al.}
\newblock \bibinfo{title}{Quantum error correction in a solid-state hybrid spin
  register}.
\newblock \emph{\bibinfo{journal}{Nature}} \textbf{\bibinfo{volume}{506}},
  \bibinfo{pages}{204--207} (\bibinfo{year}{2014}).

\bibitem{liu2017large}
\bibinfo{author}{Liu, S.} \emph{et~al.}
\newblock \bibinfo{title}{Large-scale growth of high-quality hexagonal boron
  nitride crystals at atmospheric pressure from an {Fe}--{Cr} flux}.
\newblock \emph{\bibinfo{journal}{Crystal Growth \& Design}}
  \textbf{\bibinfo{volume}{17}}, \bibinfo{pages}{4932--4935}
  (\bibinfo{year}{2017}).

\bibitem{liu2018single}
\bibinfo{author}{Liu, S.} \emph{et~al.}
\newblock \bibinfo{title}{Single crystal growth of millimeter-sized
  monoisotopic hexagonal boron nitride}.
\newblock \emph{\bibinfo{journal}{Chemistry of Materials}}
  \textbf{\bibinfo{volume}{30}}, \bibinfo{pages}{6222--6225}
  (\bibinfo{year}{2018}).

\bibitem{li2020single}
\bibinfo{author}{Li, J.} \emph{et~al.}
\newblock \bibinfo{title}{Single crystal growth of monoisotopic hexagonal boron
  nitride from a {Fe}--{Cr} flux}.
\newblock \emph{\bibinfo{journal}{Journal of Materials Chemistry C}}
  \textbf{\bibinfo{volume}{8}}, \bibinfo{pages}{9931--9935}
  (\bibinfo{year}{2020}).

\bibitem{janzen2023boron}
\bibinfo{author}{Janzen, E.} \emph{et~al.}
\newblock \bibinfo{title}{Boron and nitrogen isotope effects on hexagonal boron
  nitride properties}.
\newblock \emph{\bibinfo{journal}{Advanced Materials}} \bibinfo{pages}{2306033}
  (\bibinfo{year}{2023}).

\bibitem{gao2022nuclear}
\bibinfo{author}{Gao, X.} \emph{et~al.}
\newblock \bibinfo{title}{Nuclear spin polarization and control in hexagonal
  boron nitride}.
\newblock \emph{\bibinfo{journal}{Nature Materials}}
  \textbf{\bibinfo{volume}{21}}, \bibinfo{pages}{1024--1028}
  (\bibinfo{year}{2022}).

\bibitem{ivady2020ab}
\bibinfo{author}{Iv{\'a}dy, V.} \emph{et~al.}
\newblock \bibinfo{title}{Ab initio theory of the negatively charged boron
  vacancy qubit in hexagonal boron nitride}.
\newblock \emph{\bibinfo{journal}{npj Computational Materials}}
  \textbf{\bibinfo{volume}{6}}, \bibinfo{pages}{41} (\bibinfo{year}{2020}).

\bibitem{gottscholl2021spin}
\bibinfo{author}{Gottscholl, A.} \emph{et~al.}
\newblock \bibinfo{title}{Spin defects in hbn as promising temperature,
  pressure and magnetic field quantum sensors}.
\newblock \emph{\bibinfo{journal}{Nature Communications}}
  \textbf{\bibinfo{volume}{12}}, \bibinfo{pages}{4480} (\bibinfo{year}{2021}).

\bibitem{dreau2011avoiding}
\bibinfo{author}{Dr{\'e}au, A.} \emph{et~al.}
\newblock \bibinfo{title}{Avoiding power broadening in optically detected
  magnetic resonance of single {NV} defects for enhanced dc magnetic field
  sensitivity}.
\newblock \emph{\bibinfo{journal}{Physical Review B}}
  \textbf{\bibinfo{volume}{84}}, \bibinfo{pages}{195204}
  (\bibinfo{year}{2011}).

\bibitem{barry2020sensitivity}
\bibinfo{author}{Barry, J.~F.} \emph{et~al.}
\newblock \bibinfo{title}{Sensitivity optimization for {NV}-diamond
  magnetometry}.
\newblock \emph{\bibinfo{journal}{Reviews of Modern Physics}}
  \textbf{\bibinfo{volume}{92}}, \bibinfo{pages}{015004}
  (\bibinfo{year}{2020}).

\bibitem{yang2008quantum}
\bibinfo{author}{Yang, W.} \& \bibinfo{author}{Liu, R.-B.}
\newblock \bibinfo{title}{Quantum many-body theory of qubit decoherence in a
  finite-size spin bath}.
\newblock \emph{\bibinfo{journal}{Physical Review B}}
  \textbf{\bibinfo{volume}{78}}, \bibinfo{pages}{085315}
  (\bibinfo{year}{2008}).

\bibitem{durand2023optically}
\bibinfo{author}{Durand, A.} \emph{et~al.}
\newblock \bibinfo{title}{Optically active spin defects in few-layer thick
  hexagonal boron nitride}.
\newblock \emph{\bibinfo{journal}{Phys. Rev. Lett.}}
  \textbf{\bibinfo{volume}{131}}, \bibinfo{pages}{116902}
  (\bibinfo{year}{2023}).

\bibitem{cambria2023temperature}
\bibinfo{author}{Cambria, M.} \emph{et~al.}
\newblock \bibinfo{title}{Temperature-dependent spin-lattice relaxation of the
  nitrogen-vacancy spin triplet in diamond}.
\newblock \emph{\bibinfo{journal}{Physical Review Letters}}
  \textbf{\bibinfo{volume}{130}}, \bibinfo{pages}{256903}
  (\bibinfo{year}{2023}).

\bibitem{tang2023first}
\bibinfo{author}{Tang, H.}, \bibinfo{author}{Barr, A.~R.},
  \bibinfo{author}{Wang, G.}, \bibinfo{author}{Cappellaro, P.} \&
  \bibinfo{author}{Li, J.}
\newblock \bibinfo{title}{First-principles calculation of the
  temperature-dependent transition energies in spin defects}.
\newblock \emph{\bibinfo{journal}{The Journal of Physical Chemistry Letters}}
  \textbf{\bibinfo{volume}{14}}, \bibinfo{pages}{3266--3273}
  (\bibinfo{year}{2023}).

\bibitem{mondal2023spin}
\bibinfo{author}{Mondal, S.} \& \bibinfo{author}{Lunghi, A.}
\newblock \bibinfo{title}{Spin-phonon decoherence in solid-state paramagnetic
  defects from first principles}.
\newblock \emph{\bibinfo{journal}{npj Computational Materials}}
  \textbf{\bibinfo{volume}{9}}, \bibinfo{pages}{120} (\bibinfo{year}{2023}).

\bibitem{chatterjee2019diagnosing}
\bibinfo{author}{Chatterjee, S.}, \bibinfo{author}{Rodriguez-Nieva, J.~F.} \&
  \bibinfo{author}{Demler, E.}
\newblock \bibinfo{title}{Diagnosing phases of magnetic insulators via noise
  magnetometry with spin qubits}.
\newblock \emph{\bibinfo{journal}{Physical Review B}}
  \textbf{\bibinfo{volume}{99}}, \bibinfo{pages}{104425}
  (\bibinfo{year}{2019}).

\bibitem{chatterjee2022single}
\bibinfo{author}{Chatterjee, S.} \emph{et~al.}
\newblock \bibinfo{title}{Single-spin qubit magnetic spectroscopy of
  two-dimensional superconductivity}.
\newblock \emph{\bibinfo{journal}{Physical Review Research}}
  \textbf{\bibinfo{volume}{4}}, \bibinfo{pages}{L012001}
  (\bibinfo{year}{2022}).

\bibitem{huang2022wide}
\bibinfo{author}{Huang, M.} \emph{et~al.}
\newblock \bibinfo{title}{Wide field imaging of van der waals ferromagnet
  {$\mathrm{Fe}_3\mathrm{Ge}\mathrm{Te}_2$} by spin defects in hexagonal boron
  nitride}.
\newblock \emph{\bibinfo{journal}{Nature Communications}}
  \textbf{\bibinfo{volume}{13}}, \bibinfo{pages}{5369} (\bibinfo{year}{2022}).

\bibitem{mathur2022excited}
\bibinfo{author}{Mathur, N.} \emph{et~al.}
\newblock \bibinfo{title}{Excited-state spin-resonance spectroscopy of
  $\mathrm{V}_\mathrm{B}^-$ defects in hexagonal boron nitride}.
\newblock \emph{\bibinfo{journal}{Bulletin of the American Physical Society}}
  (\bibinfo{year}{2022}).

\bibitem{baber2021excited}
\bibinfo{author}{Baber, S.} \emph{et~al.}
\newblock \bibinfo{title}{Excited state spectroscopy of boron vacancy defects
  in hexagonal boron nitride using time-resolved optically detected magnetic
  resonance}.
\newblock \emph{\bibinfo{journal}{Nano Letters}} \textbf{\bibinfo{volume}{22}},
  \bibinfo{pages}{461--467} (\bibinfo{year}{2021}).

\bibitem{yu2022excited}
\bibinfo{author}{Yu, P.} \emph{et~al.}
\newblock \bibinfo{title}{Excited-state spectroscopy of spin defects in
  hexagonal boron nitride}.
\newblock \emph{\bibinfo{journal}{Nano Letters}} \textbf{\bibinfo{volume}{22}},
  \bibinfo{pages}{3545--3549} (\bibinfo{year}{2022}).

\bibitem{zu2014experimental}
\bibinfo{author}{Zu, C.} \emph{et~al.}
\newblock \bibinfo{title}{Experimental realization of universal geometric
  quantum gates with solid-state spins}.
\newblock \emph{\bibinfo{journal}{Nature}} \textbf{\bibinfo{volume}{514}},
  \bibinfo{pages}{72--75} (\bibinfo{year}{2014}).

\bibitem{bradley2019ten}
\bibinfo{author}{Bradley, C.~E.} \emph{et~al.}
\newblock \bibinfo{title}{A ten-qubit solid-state spin register with quantum
  memory up to one minute}.
\newblock \emph{\bibinfo{journal}{Physical Review X}}
  \textbf{\bibinfo{volume}{9}}, \bibinfo{pages}{031045} (\bibinfo{year}{2019}).

\bibitem{sangtawesin2016hyperfine}
\bibinfo{author}{Sangtawesin, S.} \emph{et~al.}
\newblock \bibinfo{title}{Hyperfine-enhanced gyromagnetic ratio of a nuclear
  spin in diamond}.
\newblock \emph{\bibinfo{journal}{New Journal of Physics}}
  \textbf{\bibinfo{volume}{18}}, \bibinfo{pages}{083016}
  (\bibinfo{year}{2016}).

\bibitem{fröch2021coupling}
\bibinfo{author}{Fröch, J.~E.} \emph{et~al.}
\newblock \bibinfo{title}{Coupling spin defects in hexagonal boron nitride to
  monolithic bullseye cavities}.
\newblock \emph{\bibinfo{journal}{Nano Letters}} \textbf{\bibinfo{volume}{21}},
  \bibinfo{pages}{6549--6555} (\bibinfo{year}{2021}).

\bibitem{mendelson2022coupling}
\bibinfo{author}{Mendelson, N.} \emph{et~al.}
\newblock \bibinfo{title}{Coupling spin defects in a layered material to
  nanoscale plasmonic cavities}.
\newblock \emph{\bibinfo{journal}{Advanced Materials}}
  \textbf{\bibinfo{volume}{34}}, \bibinfo{pages}{2106046}
  (\bibinfo{year}{2022}).

\bibitem{qian2022unveiling}
\bibinfo{author}{Qian, C.} \emph{et~al.}
\newblock \bibinfo{title}{Unveiling the zero-phonon line of the boron vacancy
  center by cavity-enhanced emission}.
\newblock \emph{\bibinfo{journal}{Nano Letters}} \textbf{\bibinfo{volume}{22}},
  \bibinfo{pages}{5137--5142} (\bibinfo{year}{2022}).

\bibitem{whitefield2023magnetic}
\bibinfo{author}{Whitefield, B.}, \bibinfo{author}{Toth, M.},
  \bibinfo{author}{Aharonovich, I.}, \bibinfo{author}{Tetienne, J.-P.} \&
  \bibinfo{author}{Kianinia, M.}
\newblock \bibinfo{title}{Magnetic field sensitivity optimization of negatively
  charged boron vacancy defects in hbn.}
\newblock \emph{\bibinfo{journal}{Advanced Quantum Technologies}}
  \bibinfo{pages}{2300118} (\bibinfo{year}{2023}).

\bibitem{zhou2023dc}
\bibinfo{author}{Zhou, F.} \emph{et~al.}
\newblock \bibinfo{title}{Dc magnetic field sensitivity optimization of spin
  defects in hexagonal boron nitride}.
\newblock \emph{\bibinfo{journal}{Nano Letters}}  (\bibinfo{year}{2023}).

\bibitem{guo2023coherent}
\bibinfo{author}{Guo, N.-J.} \emph{et~al.}
\newblock \bibinfo{title}{Coherent control of an ultrabright single spin in
  hexagonal boron nitride at room temperature}.
\newblock \emph{\bibinfo{journal}{Nature Communications}}
  \textbf{\bibinfo{volume}{14}}, \bibinfo{pages}{2893} (\bibinfo{year}{2023}).

\bibitem{stern2022room}
\bibinfo{author}{Stern, H.~L.} \emph{et~al.}
\newblock \bibinfo{title}{Room-temperature optically detected magnetic
  resonance of single defects in hexagonal boron nitride}.
\newblock \emph{\bibinfo{journal}{Nature Communications}}
  \textbf{\bibinfo{volume}{13}}, \bibinfo{pages}{618} (\bibinfo{year}{2022}).

\bibitem{li2022carbon}
\bibinfo{author}{Li, S.}, \bibinfo{author}{Thiering, G.},
  \bibinfo{author}{Udvarhelyi, P.}, \bibinfo{author}{Iv{\'a}dy, V.} \&
  \bibinfo{author}{Gali, A.}
\newblock \bibinfo{title}{Carbon defect qubit in two-dimensional
  $\mathrm{W}\mathrm{S}_2$}.
\newblock \emph{\bibinfo{journal}{Nature Communications}}
  \textbf{\bibinfo{volume}{13}}, \bibinfo{pages}{1210} (\bibinfo{year}{2022}).

\bibitem{lee2022spin}
\bibinfo{author}{Lee, Y.} \emph{et~al.}
\newblock \bibinfo{title}{Spin-defect qubits in two-dimensional transition
  metal dichalcogenides operating at telecom wavelengths}.
\newblock \emph{\bibinfo{journal}{Nature Communications}}
  \textbf{\bibinfo{volume}{13}}, \bibinfo{pages}{7501} (\bibinfo{year}{2022}).

\bibitem{jacques2023isotope}
\bibinfo{author}{Clua-Provost, T.} \emph{et~al.}
\newblock \bibinfo{title}{Isotopic control of the boron-vacancy spin defect in
  hexagonal boron nitride}.
\newblock \emph{\bibinfo{journal}{Physical Review Letters}}
  \textbf{\bibinfo{volume}{131}}, \bibinfo{pages}{126901}
  (\bibinfo{year}{2023}).

\end{thebibliography}


\begin{thebibliography}{7}%
\makeatletter
\providecommand \@ifxundefined [1]{%
 \@ifx{#1\undefined}
}%
\providecommand \@ifnum [1]{%
 \ifnum #1\expandafter \@firstoftwo
 \else \expandafter \@secondoftwo
 \fi
}%
\providecommand \@ifx [1]{%
 \ifx #1\expandafter \@firstoftwo
 \else \expandafter \@secondoftwo
 \fi
}%
\providecommand \natexlab [1]{#1}%
\providecommand \enquote  [1]{``#1''}%
\providecommand \bibnamefont  [1]{#1}%
\providecommand \bibfnamefont [1]{#1}%
\providecommand \citenamefont [1]{#1}%
\providecommand \href@noop [0]{\@secondoftwo}%
\providecommand \href [0]{\begingroup \@sanitize@url \@href}%
\providecommand \@href[1]{\@@startlink{#1}\@@href}%
\providecommand \@@href[1]{\endgroup#1\@@endlink}%
\providecommand \@sanitize@url [0]{\catcode `\\12\catcode `\$12\catcode
  `\&12\catcode `\#12\catcode `\^12\catcode `\_12\catcode `\%12\relax}%
\providecommand \@@startlink[1]{}%
\providecommand \@@endlink[0]{}%
\providecommand \url  [0]{\begingroup\@sanitize@url \@url }%
\providecommand \@url [1]{\endgroup\@href {#1}{\urlprefix }}%
\providecommand \urlprefix  [0]{URL }%
\providecommand \Eprint [0]{\href }%
\providecommand \doibase [0]{http://dx.doi.org/}%
\providecommand \selectlanguage [0]{\@gobble}%
\providecommand \bibinfo  [0]{\@secondoftwo}%
\providecommand \bibfield  [0]{\@secondoftwo}%
\providecommand \translation [1]{[#1]}%
\providecommand \BibitemOpen [0]{}%
\providecommand \bibitemStop [0]{}%
\providecommand \bibitemNoStop [0]{.\EOS\space}%
\providecommand \EOS [0]{\spacefactor3000\relax}%
\providecommand \BibitemShut  [1]{\csname bibitem#1\endcsname}%
\let\auto@bib@innerbib\@empty
\bibitem [{\citenamefont {Janzen}\ \emph {et~al.}(2023)\citenamefont {Janzen},
  \citenamefont {Schutte}, \citenamefont {Plo}, \citenamefont {Rousseau},
  \citenamefont {Michel}, \citenamefont {Desrat}, \citenamefont {Valvin},
  \citenamefont {Jacques}, \citenamefont {Cassabois}, \citenamefont {Gil},\
  and\ \citenamefont {Edgar}}]{janzen2023boron}%
  \BibitemOpen
  \bibfield  {author} {\bibinfo {author} {\bibfnamefont {E.}~\bibnamefont
  {Janzen}}, \bibinfo {author} {\bibfnamefont {H.}~\bibnamefont {Schutte}},
  \bibinfo {author} {\bibfnamefont {J.}~\bibnamefont {Plo}}, \bibinfo {author}
  {\bibfnamefont {A.}~\bibnamefont {Rousseau}}, \bibinfo {author}
  {\bibfnamefont {T.}~\bibnamefont {Michel}}, \bibinfo {author} {\bibfnamefont
  {W.}~\bibnamefont {Desrat}}, \bibinfo {author} {\bibfnamefont
  {P.}~\bibnamefont {Valvin}}, \bibinfo {author} {\bibfnamefont
  {V.}~\bibnamefont {Jacques}}, \bibinfo {author} {\bibfnamefont
  {G.}~\bibnamefont {Cassabois}}, \bibinfo {author} {\bibfnamefont
  {B.}~\bibnamefont {Gil}}, \ and\ \bibinfo {author} {\bibfnamefont {J.~H.}\
  \bibnamefont {Edgar}},\ }\href {\doibase
  https://doi.org/10.1002/adma.202306033} {\bibfield  {journal} {\bibinfo
  {journal} {Advanced Materials}\ ,\ \bibinfo {pages} {2306033}} (\bibinfo
  {year} {2023})}\BibitemShut {NoStop}%
\bibitem [{\citenamefont {Huang}\ \emph {et~al.}(2015)\citenamefont {Huang},
  \citenamefont {Sutter}, \citenamefont {Shi}, \citenamefont {Zheng},
  \citenamefont {Yang}, \citenamefont {Englund}, \citenamefont {Gao},\ and\
  \citenamefont {Sutter}}]{huang2015reliable}%
  \BibitemOpen
  \bibfield  {author} {\bibinfo {author} {\bibfnamefont {Y.}~\bibnamefont
  {Huang}}, \bibinfo {author} {\bibfnamefont {E.}~\bibnamefont {Sutter}},
  \bibinfo {author} {\bibfnamefont {N.~N.}\ \bibnamefont {Shi}}, \bibinfo
  {author} {\bibfnamefont {J.}~\bibnamefont {Zheng}}, \bibinfo {author}
  {\bibfnamefont {T.}~\bibnamefont {Yang}}, \bibinfo {author} {\bibfnamefont
  {D.}~\bibnamefont {Englund}}, \bibinfo {author} {\bibfnamefont {H.-J.}\
  \bibnamefont {Gao}}, \ and\ \bibinfo {author} {\bibfnamefont
  {P.}~\bibnamefont {Sutter}},\ }\href@noop {} {\bibfield  {journal} {\bibinfo
  {journal} {ACS Nano}\ }\textbf {\bibinfo {volume} {9}},\ \bibinfo {pages}
  {10612} (\bibinfo {year} {2015})}\BibitemShut {NoStop}%
\bibitem [{\citenamefont {Gong}\ \emph {et~al.}(2023)\citenamefont {Gong},
  \citenamefont {He}, \citenamefont {Gao}, \citenamefont {Ju}, \citenamefont
  {Liu}, \citenamefont {Ye}, \citenamefont {Henriksen}, \citenamefont {Li},\
  and\ \citenamefont {Zu}}]{gong2023coherent}%
  \BibitemOpen
  \bibfield  {author} {\bibinfo {author} {\bibfnamefont {R.}~\bibnamefont
  {Gong}}, \bibinfo {author} {\bibfnamefont {G.}~\bibnamefont {He}}, \bibinfo
  {author} {\bibfnamefont {X.}~\bibnamefont {Gao}}, \bibinfo {author}
  {\bibfnamefont {P.}~\bibnamefont {Ju}}, \bibinfo {author} {\bibfnamefont
  {Z.}~\bibnamefont {Liu}}, \bibinfo {author} {\bibfnamefont {B.}~\bibnamefont
  {Ye}}, \bibinfo {author} {\bibfnamefont {E.~A.}\ \bibnamefont {Henriksen}},
  \bibinfo {author} {\bibfnamefont {T.}~\bibnamefont {Li}}, \ and\ \bibinfo
  {author} {\bibfnamefont {C.}~\bibnamefont {Zu}},\ }\href@noop {} {\bibfield
  {journal} {\bibinfo  {journal} {Nature Communications}\ }\textbf {\bibinfo
  {volume} {14}},\ \bibinfo {pages} {3299} (\bibinfo {year}
  {2023})}\BibitemShut {NoStop}%
\bibitem [{\citenamefont {Iv{\'a}dy}\ \emph {et~al.}(2020)\citenamefont
  {Iv{\'a}dy}, \citenamefont {Barcza}, \citenamefont {Thiering}, \citenamefont
  {Li}, \citenamefont {Hamdi}, \citenamefont {Chou}, \citenamefont {Legeza},\
  and\ \citenamefont {Gali}}]{ivady2020ab}%
  \BibitemOpen
  \bibfield  {author} {\bibinfo {author} {\bibfnamefont {V.}~\bibnamefont
  {Iv{\'a}dy}}, \bibinfo {author} {\bibfnamefont {G.}~\bibnamefont {Barcza}},
  \bibinfo {author} {\bibfnamefont {G.}~\bibnamefont {Thiering}}, \bibinfo
  {author} {\bibfnamefont {S.}~\bibnamefont {Li}}, \bibinfo {author}
  {\bibfnamefont {H.}~\bibnamefont {Hamdi}}, \bibinfo {author} {\bibfnamefont
  {J.-P.}\ \bibnamefont {Chou}}, \bibinfo {author} {\bibfnamefont
  {{\"O}.}~\bibnamefont {Legeza}}, \ and\ \bibinfo {author} {\bibfnamefont
  {A.}~\bibnamefont {Gali}},\ }\href@noop {} {\bibfield  {journal} {\bibinfo
  {journal} {npj Computational Materials}\ }\textbf {\bibinfo {volume} {6}},\
  \bibinfo {pages} {41} (\bibinfo {year} {2020})}\BibitemShut {NoStop}%
\bibitem [{\citenamefont {Haykal}\ \emph {et~al.}(2022)\citenamefont {Haykal},
  \citenamefont {Tanos}, \citenamefont {Minotto}, \citenamefont {Durand},
  \citenamefont {Fabre}, \citenamefont {Li}, \citenamefont {Edgar},
  \citenamefont {Ivady}, \citenamefont {Gali}, \citenamefont {Michel} \emph
  {et~al.}}]{haykal2022decoherence}%
  \BibitemOpen
  \bibfield  {author} {\bibinfo {author} {\bibfnamefont {A.}~\bibnamefont
  {Haykal}}, \bibinfo {author} {\bibfnamefont {R.}~\bibnamefont {Tanos}},
  \bibinfo {author} {\bibfnamefont {N.}~\bibnamefont {Minotto}}, \bibinfo
  {author} {\bibfnamefont {A.}~\bibnamefont {Durand}}, \bibinfo {author}
  {\bibfnamefont {F.}~\bibnamefont {Fabre}}, \bibinfo {author} {\bibfnamefont
  {J.}~\bibnamefont {Li}}, \bibinfo {author} {\bibfnamefont {J.}~\bibnamefont
  {Edgar}}, \bibinfo {author} {\bibfnamefont {V.}~\bibnamefont {Ivady}},
  \bibinfo {author} {\bibfnamefont {A.}~\bibnamefont {Gali}}, \bibinfo {author}
  {\bibfnamefont {T.}~\bibnamefont {Michel}},  \emph {et~al.},\ }\href@noop {}
  {\bibfield  {journal} {\bibinfo  {journal} {Nature Communications}\ }\textbf
  {\bibinfo {volume} {13}},\ \bibinfo {pages} {4347} (\bibinfo {year}
  {2022})}\BibitemShut {NoStop}%
\bibitem [{\citenamefont {Reimers}\ \emph {et~al.}(2020)\citenamefont
  {Reimers}, \citenamefont {Shen}, \citenamefont {Kianinia}, \citenamefont
  {Bradac}, \citenamefont {Aharonovich}, \citenamefont {Ford},\ and\
  \citenamefont {Piecuch}}]{reimers2020photoluminescence}%
  \BibitemOpen
  \bibfield  {author} {\bibinfo {author} {\bibfnamefont {J.~R.}\ \bibnamefont
  {Reimers}}, \bibinfo {author} {\bibfnamefont {J.}~\bibnamefont {Shen}},
  \bibinfo {author} {\bibfnamefont {M.}~\bibnamefont {Kianinia}}, \bibinfo
  {author} {\bibfnamefont {C.}~\bibnamefont {Bradac}}, \bibinfo {author}
  {\bibfnamefont {I.}~\bibnamefont {Aharonovich}}, \bibinfo {author}
  {\bibfnamefont {M.~J.}\ \bibnamefont {Ford}}, \ and\ \bibinfo {author}
  {\bibfnamefont {P.}~\bibnamefont {Piecuch}},\ }\href@noop {} {\bibfield
  {journal} {\bibinfo  {journal} {Physical Review B}\ }\textbf {\bibinfo
  {volume} {102}},\ \bibinfo {pages} {144105} (\bibinfo {year}
  {2020})}\BibitemShut {NoStop}%
\bibitem [{\citenamefont {Gao}\ \emph {et~al.}(2022)\citenamefont {Gao},
  \citenamefont {Vaidya}, \citenamefont {Li}, \citenamefont {Ju}, \citenamefont
  {Jiang}, \citenamefont {Xu}, \citenamefont {Allcca}, \citenamefont {Shen},
  \citenamefont {Taniguchi}, \citenamefont {Watanabe} \emph
  {et~al.}}]{gao2022nuclear}%
  \BibitemOpen
  \bibfield  {author} {\bibinfo {author} {\bibfnamefont {X.}~\bibnamefont
  {Gao}}, \bibinfo {author} {\bibfnamefont {S.}~\bibnamefont {Vaidya}},
  \bibinfo {author} {\bibfnamefont {K.}~\bibnamefont {Li}}, \bibinfo {author}
  {\bibfnamefont {P.}~\bibnamefont {Ju}}, \bibinfo {author} {\bibfnamefont
  {B.}~\bibnamefont {Jiang}}, \bibinfo {author} {\bibfnamefont
  {Z.}~\bibnamefont {Xu}}, \bibinfo {author} {\bibfnamefont {A.~E.~L.}\
  \bibnamefont {Allcca}}, \bibinfo {author} {\bibfnamefont {K.}~\bibnamefont
  {Shen}}, \bibinfo {author} {\bibfnamefont {T.}~\bibnamefont {Taniguchi}},
  \bibinfo {author} {\bibfnamefont {K.}~\bibnamefont {Watanabe}},  \emph
  {et~al.},\ }\href@noop {} {\bibfield  {journal} {\bibinfo  {journal} {Nature
  Materials}\ }\textbf {\bibinfo {volume} {21}},\ \bibinfo {pages} {1024}
  (\bibinfo {year} {2022})}\BibitemShut {NoStop}%
\end{thebibliography}%


\begin{thebibliography}{6}%
\makeatletter
\providecommand \@ifxundefined [1]{%
 \@ifx{#1\undefined}
}%
\providecommand \@ifnum [1]{%
 \ifnum #1\expandafter \@firstoftwo
 \else \expandafter \@secondoftwo
 \fi
}%
\providecommand \@ifx [1]{%
 \ifx #1\expandafter \@firstoftwo
 \else \expandafter \@secondoftwo
 \fi
}%
\providecommand \natexlab [1]{#1}%
\providecommand \enquote  [1]{``#1''}%
\providecommand \bibnamefont  [1]{#1}%
\providecommand \bibfnamefont [1]{#1}%
\providecommand \citenamefont [1]{#1}%
\providecommand \href@noop [0]{\@secondoftwo}%
\providecommand \href [0]{\begingroup \@sanitize@url \@href}%
\providecommand \@href[1]{\@@startlink{#1}\@@href}%
\providecommand \@@href[1]{\endgroup#1\@@endlink}%
\providecommand \@sanitize@url [0]{\catcode `\\12\catcode `\$12\catcode
  `\&12\catcode `\#12\catcode `\^12\catcode `\_12\catcode `\%12\relax}%
\providecommand \@@startlink[1]{}%
\providecommand \@@endlink[0]{}%
\providecommand \url  [0]{\begingroup\@sanitize@url \@url }%
\providecommand \@url [1]{\endgroup\@href {#1}{\urlprefix }}%
\providecommand \urlprefix  [0]{URL }%
\providecommand \Eprint [0]{\href }%
\providecommand \doibase [0]{http://dx.doi.org/}%
\providecommand \selectlanguage [0]{\@gobble}%
\providecommand \bibinfo  [0]{\@secondoftwo}%
\providecommand \bibfield  [0]{\@secondoftwo}%
\providecommand \translation [1]{[#1]}%
\providecommand \BibitemOpen [0]{}%
\providecommand \bibitemStop [0]{}%
\providecommand \bibitemNoStop [0]{.\EOS\space}%
\providecommand \EOS [0]{\spacefactor3000\relax}%
\providecommand \BibitemShut  [1]{\csname bibitem#1\endcsname}%
\let\auto@bib@innerbib\@empty
\bibitem [{\citenamefont {Gong}\ \emph {et~al.}(2023)\citenamefont {Gong},
  \citenamefont {He}, \citenamefont {Gao}, \citenamefont {Ju}, \citenamefont
  {Liu}, \citenamefont {Ye}, \citenamefont {Henriksen}, \citenamefont {Li},\
  and\ \citenamefont {Zu}}]{gong2023coherent}%
  \BibitemOpen
  \bibfield  {author} {\bibinfo {author} {\bibfnamefont {R.}~\bibnamefont
  {Gong}}, \bibinfo {author} {\bibfnamefont {G.}~\bibnamefont {He}}, \bibinfo
  {author} {\bibfnamefont {X.}~\bibnamefont {Gao}}, \bibinfo {author}
  {\bibfnamefont {P.}~\bibnamefont {Ju}}, \bibinfo {author} {\bibfnamefont
  {Z.}~\bibnamefont {Liu}}, \bibinfo {author} {\bibfnamefont {B.}~\bibnamefont
  {Ye}}, \bibinfo {author} {\bibfnamefont {E.~A.}\ \bibnamefont {Henriksen}},
  \bibinfo {author} {\bibfnamefont {T.}~\bibnamefont {Li}}, \ and\ \bibinfo
  {author} {\bibfnamefont {C.}~\bibnamefont {Zu}},\ }\href@noop {} {\bibfield
  {journal} {\bibinfo  {journal} {Nature Communications}\ }\textbf {\bibinfo
  {volume} {14}},\ \bibinfo {pages} {3299} (\bibinfo {year}
  {2023})}\BibitemShut {NoStop}%
\bibitem [{\citenamefont {Dr{\'e}au}\ \emph {et~al.}(2011)\citenamefont
  {Dr{\'e}au}, \citenamefont {Lesik}, \citenamefont {Rondin}, \citenamefont
  {Spinicelli}, \citenamefont {Arcizet}, \citenamefont {Roch},\ and\
  \citenamefont {Jacques}}]{dreau2011avoiding}%
  \BibitemOpen
  \bibfield  {author} {\bibinfo {author} {\bibfnamefont {A.}~\bibnamefont
  {Dr{\'e}au}}, \bibinfo {author} {\bibfnamefont {M.}~\bibnamefont {Lesik}},
  \bibinfo {author} {\bibfnamefont {L.}~\bibnamefont {Rondin}}, \bibinfo
  {author} {\bibfnamefont {P.}~\bibnamefont {Spinicelli}}, \bibinfo {author}
  {\bibfnamefont {O.}~\bibnamefont {Arcizet}}, \bibinfo {author} {\bibfnamefont
  {J.-F.}\ \bibnamefont {Roch}}, \ and\ \bibinfo {author} {\bibfnamefont
  {V.}~\bibnamefont {Jacques}},\ }\href@noop {} {\bibfield  {journal} {\bibinfo
   {journal} {Physical Review B}\ }\textbf {\bibinfo {volume} {84}},\ \bibinfo
  {pages} {195204} (\bibinfo {year} {2011})}\BibitemShut {NoStop}%
\bibitem [{\citenamefont {Barry}\ \emph {et~al.}(2020)\citenamefont {Barry},
  \citenamefont {Schloss}, \citenamefont {Bauch}, \citenamefont {Turner},
  \citenamefont {Hart}, \citenamefont {Pham},\ and\ \citenamefont
  {Walsworth}}]{barry2020sensitivity}%
  \BibitemOpen
  \bibfield  {author} {\bibinfo {author} {\bibfnamefont {J.~F.}\ \bibnamefont
  {Barry}}, \bibinfo {author} {\bibfnamefont {J.~M.}\ \bibnamefont {Schloss}},
  \bibinfo {author} {\bibfnamefont {E.}~\bibnamefont {Bauch}}, \bibinfo
  {author} {\bibfnamefont {M.~J.}\ \bibnamefont {Turner}}, \bibinfo {author}
  {\bibfnamefont {C.~A.}\ \bibnamefont {Hart}}, \bibinfo {author}
  {\bibfnamefont {L.~M.}\ \bibnamefont {Pham}}, \ and\ \bibinfo {author}
  {\bibfnamefont {R.~L.}\ \bibnamefont {Walsworth}},\ }\href@noop {} {\bibfield
   {journal} {\bibinfo  {journal} {Reviews of Modern Physics}\ }\textbf
  {\bibinfo {volume} {92}},\ \bibinfo {pages} {015004} (\bibinfo {year}
  {2020})}\BibitemShut {NoStop}%
\bibitem [{\citenamefont {Iv{\'a}dy}\ \emph {et~al.}(2020)\citenamefont
  {Iv{\'a}dy}, \citenamefont {Barcza}, \citenamefont {Thiering}, \citenamefont
  {Li}, \citenamefont {Hamdi}, \citenamefont {Chou}, \citenamefont {Legeza},\
  and\ \citenamefont {Gali}}]{ivady2020ab}%
  \BibitemOpen
  \bibfield  {author} {\bibinfo {author} {\bibfnamefont {V.}~\bibnamefont
  {Iv{\'a}dy}}, \bibinfo {author} {\bibfnamefont {G.}~\bibnamefont {Barcza}},
  \bibinfo {author} {\bibfnamefont {G.}~\bibnamefont {Thiering}}, \bibinfo
  {author} {\bibfnamefont {S.}~\bibnamefont {Li}}, \bibinfo {author}
  {\bibfnamefont {H.}~\bibnamefont {Hamdi}}, \bibinfo {author} {\bibfnamefont
  {J.-P.}\ \bibnamefont {Chou}}, \bibinfo {author} {\bibfnamefont
  {{\"O}.}~\bibnamefont {Legeza}}, \ and\ \bibinfo {author} {\bibfnamefont
  {A.}~\bibnamefont {Gali}},\ }\href@noop {} {\bibfield  {journal} {\bibinfo
  {journal} {npj Computational Materials}\ }\textbf {\bibinfo {volume} {6}},\
  \bibinfo {pages} {41} (\bibinfo {year} {2020})}\BibitemShut {NoStop}%
\bibitem [{\citenamefont {Gao}\ \emph {et~al.}(2022)\citenamefont {Gao},
  \citenamefont {Vaidya}, \citenamefont {Li}, \citenamefont {Ju}, \citenamefont
  {Jiang}, \citenamefont {Xu}, \citenamefont {Allcca}, \citenamefont {Shen},
  \citenamefont {Taniguchi}, \citenamefont {Watanabe} \emph
  {et~al.}}]{gao2022nuclear}%
  \BibitemOpen
  \bibfield  {author} {\bibinfo {author} {\bibfnamefont {X.}~\bibnamefont
  {Gao}}, \bibinfo {author} {\bibfnamefont {S.}~\bibnamefont {Vaidya}},
  \bibinfo {author} {\bibfnamefont {K.}~\bibnamefont {Li}}, \bibinfo {author}
  {\bibfnamefont {P.}~\bibnamefont {Ju}}, \bibinfo {author} {\bibfnamefont
  {B.}~\bibnamefont {Jiang}}, \bibinfo {author} {\bibfnamefont
  {Z.}~\bibnamefont {Xu}}, \bibinfo {author} {\bibfnamefont {A.~E.~L.}\
  \bibnamefont {Allcca}}, \bibinfo {author} {\bibfnamefont {K.}~\bibnamefont
  {Shen}}, \bibinfo {author} {\bibfnamefont {T.}~\bibnamefont {Taniguchi}},
  \bibinfo {author} {\bibfnamefont {K.}~\bibnamefont {Watanabe}},  \emph
  {et~al.},\ }\href@noop {} {\bibfield  {journal} {\bibinfo  {journal} {Nature
  Materials}\ }\textbf {\bibinfo {volume} {21}},\ \bibinfo {pages} {1024}
  (\bibinfo {year} {2022})}\BibitemShut {NoStop}%
\bibitem [{\citenamefont {Puebla}\ \emph {et~al.}(2022)\citenamefont {Puebla},
  \citenamefont {Li}, \citenamefont {Zhang},\ and\ \citenamefont
  {Castellanos-Gomez}}]{Puebla2022color}%
  \BibitemOpen
  \bibfield  {author} {\bibinfo {author} {\bibfnamefont {S.}~\bibnamefont
  {Puebla}}, \bibinfo {author} {\bibfnamefont {H.}~\bibnamefont {Li}}, \bibinfo
  {author} {\bibfnamefont {H.}~\bibnamefont {Zhang}}, \ and\ \bibinfo {author}
  {\bibfnamefont {A.}~\bibnamefont {Castellanos-Gomez}},\ }\href@noop {}
  {\bibfield  {journal} {\bibinfo  {journal} {Advanced Photonics Research}\
  }\textbf {\bibinfo {volume} {3}},\ \bibinfo {pages} {2100221} (\bibinfo
  {year} {2022})}\BibitemShut {NoStop}%
\end{thebibliography}%

\end{document}


\title{Supplementary Information and Extended Figures: \texorpdfstring{\\}{} Isotope Engineering for Spin Defects in van der Waals Materials}

\author{
Ruotian~Gong,$^{1,*}$ 
Xinyi~Du,$^{1,*}$
Eli~Janzen, $^{2}$
Vincent~Liu,$^{3}$
Zhongyuan~Liu,$^{1}$
Guanghui~He,$^{1}$
Bingtian~Ye,$^{3}$ \\
Tongcang~Li,$^{4,5}$
Norman~Y.~Yao,$^{3}$
James~H.~Edgar,$^{2}$
Erik~A.~Henriksen,$^{1,6}$ 
Chong~Zu$^{1,6,\dag}$
\\
\medskip
\normalsize{$^{1}$Department of Physics, Washington University, St. Louis, MO 63130, USA}\\
\normalsize{$^{2}$Tim Taylor Department of Chemical Engineering,
Kansas State University, Manhattan, KS 66506, USA}\\
\normalsize{$^{3}$Department of Physics, Harvard University, Cambridge, MA 02138, USA}\\
\normalsize{$^{4}$Department of Physics and Astronomy, Purdue University, West Lafayette, Indiana 47907, USA}\\
\normalsize{$^{5}$Elmore Family School of Electrical and Computer Engineering, Purdue University, West Lafayette, IN 47907, USA}\\
\normalsize{$^{6}$Institute of Materials Science and Engineering, Washington University, St. Louis, MO 63130, USA}\\
\normalsize{$^*$These authors contributed equally to this work}\\
\normalsize{$^\dag$To whom correspondence should be addressed; E-mail: zu@wustl.edu}\\
}

\date{\today}

\maketitle

\tableofcontents

\section{Experimental Setup}

We characterize the spin properties of \vbm ensembles in both \hbn and \hbnnat using a homebuilt confocal laser microscope. A $532~$nm laser (Millennia eV High Power CW DPSS Laser) is used for both \vbm spin initialization and detection. The laser is shuttered by an acousto-optic modulator (AOM, G$\&$H AOMO 3110-120) in a double-pass configuration to achieve $>10^5:1$ on/off ratio. An objective lens (Mitutoyo Plan Apo 100x 378-806-3) focuses the laser beam to a diffraction-limited spot with diameter $\sim 0.6~\mu$m and collects the \vbm fluorescence. The fluorescence is then separated from the laser beam by a dichroic mirror and filtered through a long-pass filter before being detected by a single photon counting module (Excelitas SPCM-AQRH-63-FC). The signal is then processed by a data acquisition device (National Instruments USB-6343). The objective lens is mounted on a piezo objective scanner (Physik Instrumente PD72Z1x PIFOC), which controls the position of the objective and scans the laser beam vertically. The lateral scanning is performed by an X-Y galvanometer (Thorlabs GVS212).

To isolate an effective two-level system {$|m_s = 0, -1\rangle$ or $|m_s = 0, +1\rangle$},  we position a  cylindrical N40 Neodymium permanent magnet with diameter 0.750" (19.05mm) and length 1" (25.40mm) directly on top of the sample to create an strong external magnetic field along the c-axis of the hBN lattice.
Under this magnetic field, the $|m_s = \pm1\rangle$ sublevels of the \vbm are separated due to the Zeeman effect and exhibit a splitting $2\gamma_e B$, where $\gamma_e = (2\pi)\times 2.8~$MHz/G is the gyromagnetic ratio of the \vbm electronic spin.
A resonant microwave drive is applied to address the transition between the two electronic spin sublevels, whose frequency depends on the strength of the external magnetic field and is probed via the electron spin resonance (ESR) measurement.
Using a translation stage, we are also able to adjust the alignment angle of the magnetic field to explore its effect on nuclear spin polarization.
We first move the $\hat{z}$ axis of the translation stage to approach the required external magnetic field strength $B\sim760~$G for esLAC and then carefully adjust the $\hat{x}$ and $\hat{y}$ axes of the translation stage so that the nuclear polarization is maximized. 
We investigate the relationship between the alignment of $B$ and the nuclear polarization by walking the $\hat{x}$ axis of the translation stage incrementally and recording the corresponding ESR signal to characterize polarization.
By simulating the magnetic field vector of the specific magnet we use, we are able to map the translation distance to the alignment angle assuming the starting maximum point corresponds to $0^\circ$.

The microwave driving field is generated by mixing the output from a microwave source (Stanford Research SG386) and an arbitrary wave generator (AWG, Chase Scientific Wavepond DAx22000). 
The AWG we use has a sampling rate of $2~$GHz ($0.5~$ns temporal resolution), sufficiently fast to generate high-fidelity pulses to control the spin state of \vbm ensembles.
Specifically, a high-frequency signal at $ (2\pi)\times 1$ GHz to $ (2\pi)\times 6$ GHz from the microwave source is combined with a $ (2\pi)\times 0.125~$GHz signal from the AWG using a built-in in-phase/quadrature (IQ) modulator so that the sum frequency is resonant with one of the $|m_s=0\rangle \Longleftrightarrow |m_s=\pm1\rangle$ transitions. 
By modulating the amplitude, duration, and phase of the AWG output, we can control the strength, rotation angle, and axis of the microwave pulses.
For radio-frequency driving fields, we use another signal generator (Stanford Research SG384).
Both the RF and microwave signals are shuttered by a switch (Minicircuits ZASWA-2-50DRA+) to prevent any leakage. 
The microwave and RF signal are amplified by amplifiers (Mini-Circuits ZHL-15W-422-S+ for microwave and LZY-22+ for RF) and delivered to the hBN sample through a coplanar waveguide.
For nuclear resonance measurements, we combine the two signals using a 2 way power splitter (Mini-Circuits ZAPD-30-S+).
The equipment is gated through a programmable multi-channel pulse generator (SpinCore PulseBlasterESR-PRO 500) with a $2$~ns temporal resolution.

To investigate the temperature dependence of the spin relaxation timescales, we load both \hbn and \hbnnat sample into an optical accessible low-vibration cryostat (Four Nine Design SideKick Cryogenic Systems).
We use a PID cryogenic temperature controller (Lake Shore Model 336) that operates from $4~$K up to $350~$K for the temperature control.

\section{The Fourier Transformation Method for Calculating ESR Spectra}

To further illustrate the methodology used in obtaining the simulated ESR spectra for hBN samples with different isotope choices (Main text Fig.~1c and Methods Fig.~S3), we detail the calculation of spectral density, $S(\omega)$, which describes the density of states in the energy spectrum. We define $S(\omega)$ as a sum of Dirac functions over all possible $m_{\mathcal{I}}$, each centered at the respective energy $\sum_j A^j_{zz} m^j_{\mathcal{I}}$, where $m_{\mathcal{I}}^j$ represents the nuclear spin magnetic quantum number of the $j$th nuclear spin $\mathcal{I}^j$. This can be represented as

\begin{equation}
S(\omega)=\sum_{\mathrm{all\ possible\ }\{m_{\mathcal{I}} \}} \delta(\omega-\sum_j A^j_{zz} m^j_{\mathcal{I}}).
\end{equation}

We then establish the Fourier transform and its inverse as

\begin{equation}
\tilde{f}(\omega)=\mathcal{F}[f(t)]= \int_{-\infty}^{+\infty}f(t)e^{i\omega t}\mathrm{d} t
\end{equation}

\begin{equation}
f(t)=\mathcal{F}^{-1}[\tilde{f}(\omega)]= \frac{1}{2\pi} \int_{-\infty}^{+\infty}\tilde{f}(\omega)e^{-i\omega t}\mathrm{d} \omega.
\end{equation}

We can rewrite $S(\omega)$ in terms of the Fourier transform as

\begin{equation} \label{eq1}
\begin{aligned}
S(\omega) &= \sum_{\mathrm{all\ possible\ }\{m_{\mathcal{I}} \}} \frac{1}{2\pi} \mathcal{F}\{e^{- i \sum_{j}A_{zz}^j m^j_{\mathcal{I}}t}\} \\
&= \sum_{\mathrm{all\ possible\ }\{m_{\mathcal{I}} \}} \frac{1}{2\pi} \mathcal{F}\{\prod_j e^{- i A_{zz}^j m^j_{\mathcal{I}}t}\} \\
&=\frac{1}{2\pi}\mathcal{F}\{\prod_j \sum_{-\mathcal{I}^j\leq m^j_{\mathcal{I}}\leq \mathcal{I}^j} e^{- i A_{zz}^j m^j_{\mathcal{I}}t}\}.
\end{aligned}
\end{equation}

By employing these techniques, we circumvent computationally expensive calculations involving massive summations, replacing them with a tolerable number of steps involving summations, sequential multiplications, and the Fast Fourier Transform (FFT).

\section{Differential Measurement Scheme} \label{SectionDiff}

\begin{figure}[t]
    \centering
    \includegraphics[width=0.7\textwidth]{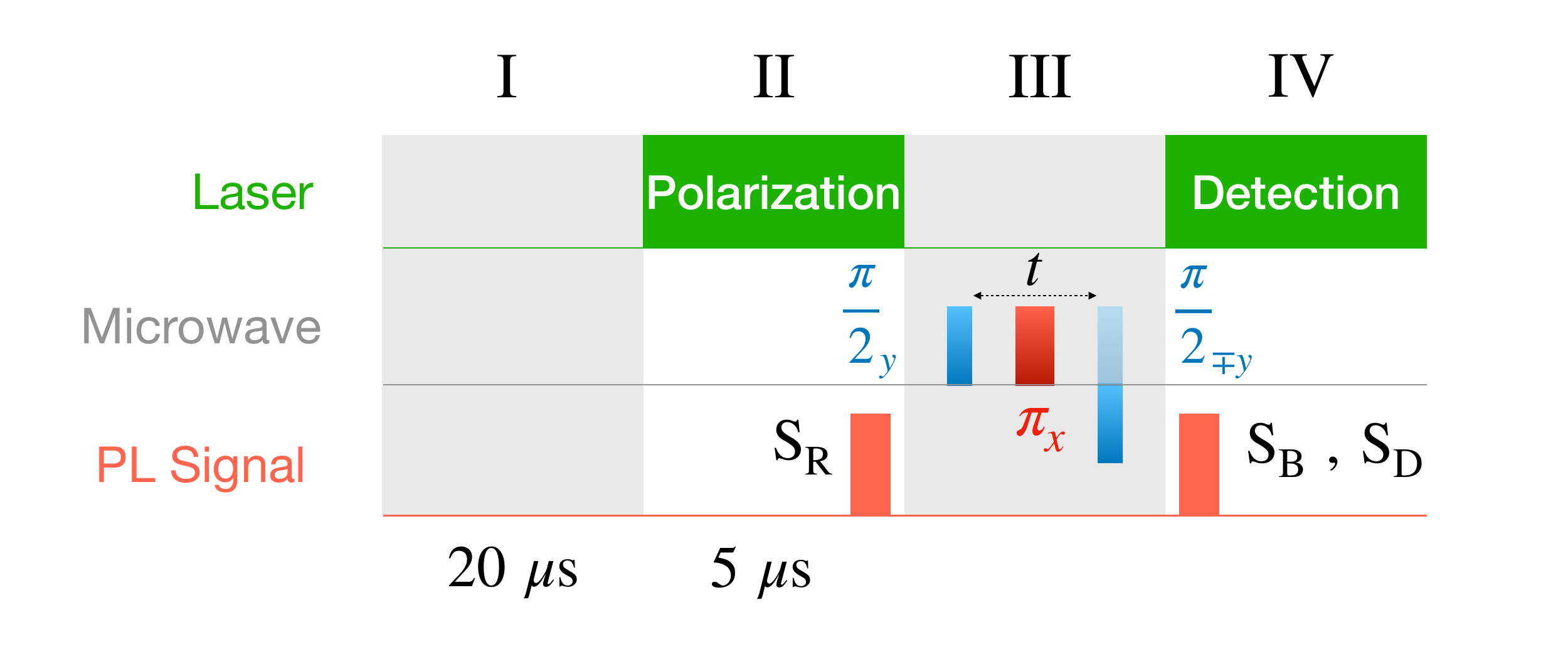}
    \caption{ {\bf Differential Measurement} Differential measurement sequence for spin echo. $\mathrm{I}$: $20~\mu$s wait time to reach charge state equilibration. $\mathrm{II}$: $5~\mu$s laser pulse to initialize the \vbm spin to $|m_s = 0\rangle$, with the reference signal, $\mathrm{S}_\mathrm{R}(t)$, collected at the end of the laser pulse. $\mathrm{III}$: microwave wave pulses for spin echo measurement; for the bright signal, a final $\frac{\pi}{2}$ pulse along the $-\hat{y}$ axis is applied; while for the dark signal, a final $\frac{\pi}{2}$ pulse along the $+\hat{y}$ axis is applied to rotate the spin to an orthogonal state. $\mathrm{IV}$: laser pulse to detect the spin state.}
    \label{Supp_Diff}
\end{figure}

To accurately probe the spin dynamics of \vbm, we utilize a robust differential measurement scheme illustrated in Figure~\ref{Supp_Diff} \cite{gong2023coherent}. 
Specifically, after letting the spin system reach charge state equilibration for $20~\mu$s without any laser illumination (I), we apply a $5~\mu$s laser pulse ($532~$nm) to initialize the spin state of \vbm (II), followed by the measurement pulse sequences (III).
Taking spin echo coherent measurements on the $|0\rangle$ and $|-1\rangle$ state as an example, we first apply a $\frac{\pi}{2}$-pulse along the $\hat{y}$ axis to prepare the system in a superposition state $\otimes_{i} \frac{|0\rangle_i+|-1\rangle_i}{\sqrt{2}}$, and then let it evolve for time $t$.
A refocusing $\pi$-pulse along the $\hat{x}$ axis at time $t/2$ is used to decouple the spin ensemble from static magnetic noise.
A final $\frac{\pi}{2}$-pulse along the $-\hat{y}$ direction rotates the spin back to the $\hat{z}$ axis for fluorescence detection (IV) and the measured photon count is designated as the bright signal $\mathrm{S}_\mathrm{B}(t)$.
By repeating the same sequence but with a final $\frac{\pi}{2}$-pulse along the positive $+\hat{y}$ axis before readout, we measure
the fluorescence of an orthogonal spin state to be the dark signal $\mathrm{S}_\mathrm{D}(t)$.
The difference between the two measurements $\mathrm{C}(t) \equiv [\mathrm{S}_\mathrm{B}(t)-\mathrm{S}_\mathrm{D}(t)]/\mathrm{S}_\mathrm{R}(t)$ can faithfully represent the measured spin coherent dynamics of \vbm, where $\mathrm{S}_\mathrm{R}(t)$ is a reference signal we measure at the end of the initialization laser pulse (II).

\section{Magnetic Field Sensitivity Estimate}

\begin{figure}[t]
    \centering
    \includegraphics[width=0.8\textwidth]{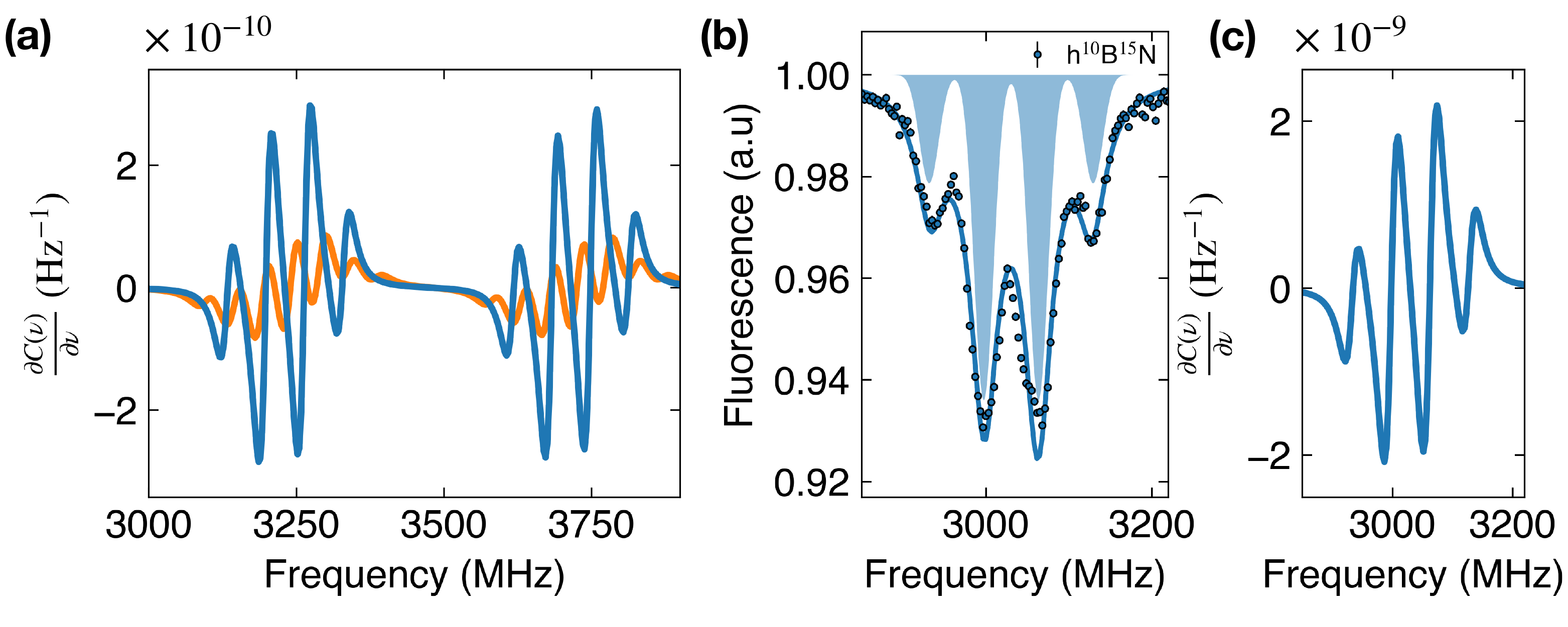}
    \caption{{\bf Optimization of magnetic field sensitivity for ESR measurements}
    (a) Derivatives of the ESR spectra of \vbm in \hbn and \hbnnat from main text Figure~1c, showing slopes with maximum at $8.2 \times 10^{-11} ~\mathrm{Hz}^{-1}$ and $3.0\times 10^{-10}~\mathrm{Hz}^{-1}$ respectively.
    (b) ESR spectrum of \vbm in \hbn after optimizing the laser and microwave powers for better sensitivity.
    (c) Derivatives of the fitted ESR spectrum in (b).}
    \label{Supp_sensitivity}
\end{figure}

The static magnetic field sensitivity of ESR measurements takes the form \cite{dreau2011avoiding,barry2020sensitivity}

\begin{equation} \label{eqDCSensitivity}
\begin{split}
\eta_\mathrm{DC} \approx\frac{2\pi}{\gamma_e \sqrt{R}} (\max|\frac{\partial{C(\nu)}}{\partial{\nu}}|)^{-1} \approx \frac{8\pi}{3\sqrt{3}} \frac{1}{\gamma_e}\frac{\Delta \nu} {C_m\sqrt{R}},
\end{split}
\end{equation} 
where $\gamma_e$ denotes the electron gyromagnetic ratio ($2\pi\times 2.8~$MHz/G), $R$ the photon detection rate, $C(\nu)$ the ESR measurement contrast at microwave frequency $\nu$, $C_m$ the maximum contrast, and $\Delta \nu$ the FWHM linewidth assuming a single Lorenztian resonance. 
If we directly compare the fitted FWHM from \vbm in \hbn and \hbnnat (main text Fig.~1c), we obtain a factor of $\sim 1.8$ improvement in sensitivity.
However, directly comparing the maximum $\frac{\partial{C(\nu)}}{\partial{\nu}}$ of each spectrum more accurately reflects the sensitivity as the hyperfine levels largely overlap with each other in \hbnnat and one cannot resolve individual Lorenztian.
In Figure~\ref{Supp_sensitivity}a, we plot $\frac{\partial{C(\nu)}}{\partial{\nu}}$ against frequency for the corresponding ESR spectrum of \vbm in \hbn and \hbnnat from main text Figure~1, where the curves are obtained by differentiating the fitted Lorentzians.
%
Here, we find the steepest slopes are $8.2 \times 10^{-11}~\mathrm{Hz}^{-1}$ and $3.0\times 10^{-10}~\mathrm{Hz}^{-1}$ for \hbnnat and \hbn respectively, and after accounting their similar photon detection rate and ESR contrast, the sensitivity enhancement increases to $\sim$ 4-fold.

We remark that, when performing the ESR measurement in main text Figure~1c, we utilize small laser and microwave powers to avoid spectral broadening to resolve the intrinsic linewidth of the resonances. 
However, to optimize sensitivity, we may want to further fine-tune the laser and microwave powers.
Specifically, increasing laser power leads to a higher count rate $R$ but lower maximum contrast and broadened linewidth, while increasing microwave power can boost contrast but also cause power broadening of the resonances.
By carefully adjusting the laser and microwave power while monitoring the ESR signal, we obtain the an improved spectrum for sensing (Fig.~\ref{Supp_sensitivity}b).
In Figure~\ref{Supp_sensitivity}c, we apply the same differentiation process to the new ESR spectrum, achieving a DC magnetic field sensitivity of $\sim 10~\mu$T~$\mathrm{Hz}^{-\frac{1}{2}}$ (with count rate $R\approx 2.7\times10^6$ photons per second).
%
%

For AC magnetic fields, the sensitivity is given by \cite{barry2020sensitivity} 
\begin{equation} \label{eqACSensitivity}
\begin{split}
\eta_\mathrm{AC} \approx \frac{\pi}{2\gamma_e} \dfrac{1}{C_\mathrm{max} e^{-(\tau/T_2)} \sqrt{\mathcal{N}}} \frac{\sqrt{t_I + \tau + t_R}}{\tau},
\end{split}
\end{equation} 
where $C_\mathrm{max}$ is the maximum $T_2$ measurement contrast, $\mathcal{N}$ is the average photon count collected in a single experimental sequence, $\tau$ is the full field interrogation time, and $t_I$ and $t_R$ are respectively the initialization and readout times.
In our experiment, a minimum of $t_I + t_R \approx 2~\mu$s is required to optically polarize and detect \vbm, which is much larger than the measurement time $\tau = T_2$; in this limit, $\eta_\mathrm{AC} \propto 1/T_2$.

\begin{figure}
    \centering 
    \includegraphics[width = 0.7\textwidth]{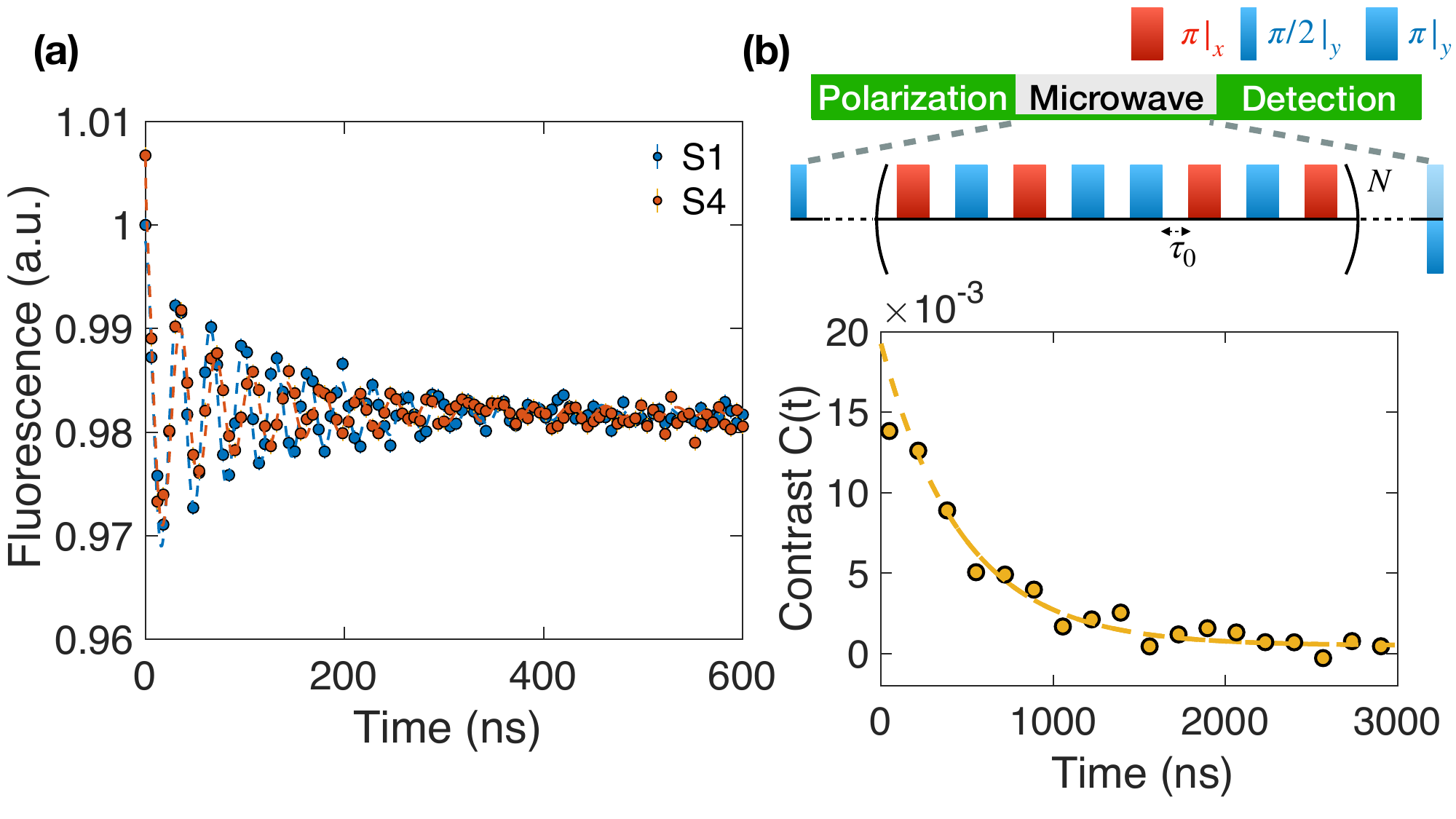}
    \caption{{\bf Rabi Oscillations and XY8 $T_2$ Measurement} (a) Rabi oscillations recorded on \hbn sample S1 and \hbnnat sample S4 under the same microwave power. 
    (b) Coherent measurement of \vbm in \hbn sample S3 using the XY8 pulse sequence. The sequence repeats itself every 8 pulses, and we take a measurement every 4 pulses to increase the number of data points. The $\pi$-pulse length $t_\pi$ is fixed at $24~$ns with the interval between every adjacent pulse fixed at $\tau_0 =18~$ns. The coherence decay is measured via increasing the number of decoupling pulses $N$.}
    \label{fig:Supp_RabiNXY8}
\end{figure}

To perform $T_2$ measurements, we first record the Rabi oscillations for pulse control. Figure~\ref{fig:Supp_RabiNXY8}a shows the Rabi signal of \vbm in \hbn sample S1 when driving one of the two center resonances and in \hbnnat sample S4 when driving the center resonance.
The spin echo measurement is described in the previous section \ref{SectionDiff}.
To further improve the AC magnetic field sensitivity, we utilize a more advanced dynamical decoupling sequence, XY8, to extend the $T_2$ coherence time.
Instead of a single refocusing $\pi$-pulse, XY8 employs a series of $\pi$-pulses with alternating phases to better decouple the \vbm from the nearby nuclear spin bath (Fig.~\ref{fig:Supp_RabiNXY8}b).
The measured XY8 coherence time $T_2^\mathrm{XY} \approx 501~$ns is more than two times the spin echo coherence time $T_2^\mathrm{E}\approx 200~$ns.
Using the measured contrast, $C \approx 2\%$, and average count per sequence, $\mathcal{N} \approx 0.27$, where we estimate that the AC magnetic field sensitivity $\eta_\mathrm{AC}$ is optimized with a value $\approx 7    ~\mu$T~$\mathrm{Hz}^{-\frac{1}{2}}$.
%
The AC magnetic field sensing frequency is set by the corresponding filter function of the pulse sequence in the frequency domain \cite{barry2020sensitivity}. For our measurement of XY8 with $\pi$-pulse length $=24~$ns and pulse interval $=18~$ns, the frequency of the detected AC signal is $\sim 12~$MHz.

\section{\texorpdfstring{Interaction Hamiltonian for $\mathrm{V}_\mathrm{B}^-$}{} and nearby Nuclear Spins}

In this section, we derive the Hamiltonian governing the interaction between the \vbm electronic spin and nearby nuclear spins. 
The \vbm electronic spin-1 operators can be written as
\begin{equation} \label{eq7}
\begin{split}
S_z =
\begin{bmatrix}
1 & 0 & 0\\
0 & 0 & 0\\
0 & 0 & -1
\end{bmatrix}	,~
S_x = \frac{1}{\sqrt{2}}
\begin{bmatrix}
0 & 1 & 0\\
1 & 0 & 1\\
0 & 1 & 0
\end{bmatrix}	,~
S_y = \frac{1}{\sqrt{2}i}
\begin{bmatrix}
0 & 1 & 0\\
-1 & 0 & 1\\
0 & -1 & 0
\end{bmatrix}   .
\end{split}
\end{equation}
We can also define the spin raising and lowering operators for the \vbm electron spin as
\begin{equation} \label{eq8}
S_+ = \sqrt{2}
\begin{bmatrix}
0 & 1 & 0\\
0 & 0 & 1\\
0 & 0 & 0
\end{bmatrix}
= S_x + iS_y ,~
S_- = \sqrt{2}
\begin{bmatrix} 
0 & 0 & 0\\
1 & 0 & 0\\
0 & 1 & 0
\end{bmatrix}
= S_x - iS_y,
\end{equation}
and rewrite \vbm spin operators in terms of the raising and lowering operators in the form
\begin{equation} \label{eq9}
S_x = \frac{S_+ + S_-}{2} ,~
S_y = \frac{S_+ - S_-}{2i} .
\end{equation}

For nuclear spins, we consider spin-1/2 \nfive nuclei with spin operators
\begin{equation} \label{eq10}
\begin{split}
I_z = \frac{1}{2}
\begin{bmatrix}
1 & 0 \\
0 & -1
\end{bmatrix}	,~
I_x = \frac{1}{2}
\begin{bmatrix}
0 & 1 \\
1 & 0
\end{bmatrix}	,~
I_y = \frac{1}{2i}
\begin{bmatrix}
0 & 1 \\
-1 & 0
\end{bmatrix}   .
\end{split}
\end{equation}
and spin ladder operators
\begin{equation} \label{eq11}
I_+ =
\begin{bmatrix}
0 & 1 \\
0 & 0
\end{bmatrix}
= I_x + iI_y ,~
I_- =
\begin{bmatrix}
0 & 0 \\
1 & 0
\end{bmatrix}
= I_x - iI_y ,
\end{equation}
Similarly, we can rewrite \nfive nuclear spin operators as
\begin{equation} \label{eq12}
I_x = \frac{I_+ + I_-}{2} ,~
I_y = \frac{I_+ - I_-}{2i} .
\end{equation}

To expand the hyperfine interaction term between \vbm and three \nfive nuclear spins --- $\sum_{j=1}^{3}\mathbf{S}\mathbf{A}^{j}\mathbf{I}^j$ --- we note the hyperfine parameters tensor $\mathbf{A}$ for a single nuclear spin takes the form
\begin{equation} \label{eq13}
\mathbf{A} =
\begin{bmatrix}
A_{xx} & A_{xy} & A_{xz} \\
A_{yx} & A_{yy} & A_{yz} \\
A_{zx} & A_{zy} & A_{zz}
\end{bmatrix}
=
\begin{bmatrix}
A_{xx} & A_{xy} & 0 \\
A_{yx} & A_{yy} & 0 \\
0      & 0      & A_{zz}
\end{bmatrix}.
\end{equation}
Here the $\hat{z}$-axis is defined along the c-axis of hBN (perpendicular to the lattice plane, see main text Figure.~1), $\hat{x}$ and $\hat{y}$ lie in the lattice plane, with $\hat{x}$ oriented along one of the three in-plane nitrogen bonds. Due to the mirror symmetry of \vbm with respect to the $\hat{x}-\hat{y}$ plane, the four terms $A_{xz}$ $ A_{yz}$ $ A_{zx}$ $ A_{zy}$ vanish \cite{ivady2020ab, gao2022nuclear}. We can then expand the hyperfine interacting Hamiltonian to its full form
\begin{equation} \label{eq14}
\begin{split}
    \sum_{j=1}^{3}\mathbf{S}\mathbf{A}^{j}\mathbf{I}^j & = \sum_{j=1}^{3} (A_{zz}^{j}S_z I_z^j + A_{xx}^{j}S_x I_x^j +A_{yy}^{j}S_y I_y^j +A_{xy}^{j}S_xI_y^j +A_{yx}^{j}S_yI_x^j ) \\
    & = \sum_{j=1}^{3} [A_{zz}^{j}S_z I_z^j + \frac{1}{4}A_{xx}^{j}(S_+ + S_-)(I_+^j+I_-^j)-\frac{1}{4i}A_{yy}^{j}(S_+ - S_-) (I_+^j-I_-^j) \\
    & ~~~ + \frac{1}{4i}A_{xy}^{j}(S_+ + S_-) (I_+^j-I_-^j) + \frac{1}{4i}A_{yx}^{j}(S_+ - S_-) (I_+^j+I_-^j)] \\ 
    & = \sum_{j=1}^{3} [A_{zz}^{j} S_z I_z^j + \dfrac{A_{xx}+A_{yy}}{4} (S_+I_-^j  + S_-I_+^j) \\
    & ~~~ +  (\dfrac{A_{xx}-A_{yy}}{4} + \dfrac{A_{xy}}{2i}) S_+I_+^j + (\dfrac{A_{xx}-A_{yy}}{4} - \dfrac{A_{xy}}{2i}) S_-I_-^j].
\end{split}
\end{equation}
In the last step, we also use the fact that $A_{xy} = A_{yx}$ from the symmetry \cite{ivady2020ab}. Finally, since $(S_+I_-^j)^\dag = (S_-I_+^j)$ and $(S_-I_-^j)^\dag = (S_-I_-^j)$, we can further simplify the expression to the form
\begin{equation} \label{eq15}
    \sum_{j=1}^{3}\mathbf{S}\mathbf{A}^{j}\mathbf{I}^j = \sum_{j=1}^{3} [A_{zz}^{j}S_z I_z^j + (A_1^{j}S_+ I_-^j + h.c.) + (A_2^{j}S_+ I_+^j + h.c.)],
\end{equation}
where we define $A_1^j = \frac{1}{4}(A_{xx}^j+A_{yy}^j)$ and $A_2^j = \frac{1}{4} (A_{xx}^j-A_{yy}^j)+\frac{1}{2i}A_{xy}^j$.

\section{Enhanced Gyromagnetic Ratio for nuclear spin driving}

\subsection{Derivation of effective nuclear gyromagnetic ratio}\label{derivationGamma_n}

In this section, we derive the enhanced effective gyromagnetic ratio of nuclear spins in the \vbm center. For each nearest-neighbour \nfive nuclear spin, the hyperfine Hamiltonian is described by Eqn.~\ref{eq15},
where by threefold rotational symmetry, 
$|A_1^j|$ and $|A_2^j|$ are the same for each nucleus. 
%
%
We note that this form does not hold in the excited state due to the breaking of threefold rotational symmetry by the Jahn-Teller effect.

To obtain the effective gyromagnetic ratio, it is instructive to first focus on a simplified case where the electronic spin couples to only one nuclear spin. The ground-state Hamiltonian can be written as a sum of secular and non-secular terms, denoted $H_0$ and $\delta H$ respectively, with 
\begin{subequations}
\begin{align}
    H_0 &= D_{gs} S_z^2 + \gamma_e B_z S_z - \gamma_n B_z I_z + A_{zz} S_z I_z, \\
    \delta H &= \gamma_e B_\mathrm{dr} \left( \cos \theta S_x + \sin \theta S_y \right) - \gamma_n B_\mathrm{dr} \left( \cos \theta I_x + \sin \theta I_y \right) + \left( A_1 S_+ I_- + h.c. \right) + \left( A_2 S_+ I_+ + h.c. \right) \nonumber \\
    &= \frac{\gamma_e}{2} B_\mathrm{dr} \left( e^{i \theta} S_+ + e^{-i \theta} S_- \right) - \frac{\gamma_n}{2} B_\mathrm{dr} \left( e^{i \theta} I_+ + e^{-i \theta} I_- \right) + \left( A_1 S_+ I_- + h.c. \right) + \left( A_2 S_+ I_+ + h.c. \right),
\end{align}
\end{subequations}
where we have included microwave and RF drives with an transverse magnetic field $B_\mathrm{dr}$ at an angle $\theta$ to the $\hat{x}$ axis. 
Throughout our experiment, we always work at a magnetic field far away from ground-state anti-crossing (gsLAC, $B_z \approx 1240~$G), so that $\delta H$ is suppressed due to the large splitting between the $|m_s=0\rangle$ and $|m_s=\pm1\rangle$ electronic spin levels of \vbm.
In this case, we can treat $\delta H$ as a perturbation to $H_0$. Performing second-order perturbation theory, we find that the coupling term between nuclear spin sublevels can be written as
\begin{equation}
    -\frac{\gamma_n}{2} B_\mathrm{dr} \left( e^{i \theta} I_+ + e^{-i \theta} I_- \right) - \frac{\gamma_e}{D_{gs}\pm\gamma_e B_z} B_\mathrm{dr} \left[ \left(A_1 e^{i \theta} + A_2^* e^{-i \theta} \right) I_- + h.c. \right]
\end{equation}
for $|m_s = \pm 1\rangle$ states, and 
\begin{equation}
    -\frac{\gamma_n}{2} B_\mathrm{dr} \left( e^{i \theta} I_+ + e^{-i \theta} I_- \right) - \gamma_e \left( \frac{1}{D_{gs} + \gamma_e B_z} + \frac{1}{D_{gs} - \gamma_e B_z} \right) B_\mathrm{dr} \left[ \left(A_1 e^{i \theta} + A_2^* e^{-i \theta} \right) I_- + h.c.\right]
\end{equation}
for the $|m_s = 0\rangle$ state.
Focusing on the $|m_s = -1\rangle$ level at esLAC where we perform the nuclear spin control and noting that $\gamma_n \ll \gamma_e \frac{\left| A_{1,2} \right|}{D_{gs}-\gamma_e B_z}$, we drop the first term and find that the effective nuclear spin Rabi frequency is $2\gamma_e B_\mathrm{dr} \frac{\left| A_1 e^{i \theta} + A_2^* e^{-i \theta} \right|}{D_{gs}-\gamma_e B_z}$, with a nuclear gyromagnetic ratio enhancement factor of $2 \frac{\gamma_e}{\gamma_n} \frac{\left| A_1 e^{i \theta} + A_2^* e^{-i \theta} \right|}{D_{gs}-\gamma_e B_z}$.

In the \vbm center, there are three nearest-neighbor nuclear spins instead of one. However, the perturbation theory analysis above is still valid for each individual nuclear spin because there are no terms that directly couple the nuclear spins of different nuclei. Thus, we can write the nuclear coupling terms in the $\left| m_s=-1, m_I^1, m_I^2, m_I^3 \right>$ basis as the matrix
\begin{equation}
\begin{bmatrix}
0 & \omega_3 & \omega_2 & 0 & \omega_1 & 0 & 0 & 0 \\
\omega_3^* & 0 & 0 & \omega_2 & 0 & \omega_1 & 0 & 0 \\
\omega_2^* & 0 & 0 & \omega_3 & 0 & 0 & \omega_1 & 0 \\
0 & \omega_2^* & \omega_3^* & 0 & 0 & 0 & 0 & \omega_1 \\
\omega_1^* & 0 & 0 & 0 & 0 & \omega_3 & \omega_2 & 0 \\
0 & \omega_1^* & 0 & 0 & \omega_3^* & 0 & 0 & \omega_2 \\
0 & 0 & \omega_1^* & 0 & \omega_2^* & 0 & 0 & \omega_3 \\
0 & 0 & 0 & \omega_1^* & 0 & \omega_2^* & \omega_3^* & 0
\end{bmatrix},
\end{equation}
where $\omega_j \equiv -\gamma_e B_\mathrm{dr} \frac{A_1^j e^{i \theta} + \left(A_2^j\right)^* e^{-i \theta}}{D_{gs}-\gamma_e B_z}$ corresponds to the $j$-th nuclear spin. The eigenvalues of this matrix are $\pm \left| \omega_1 \right| \pm \left| \omega_2 \right| \pm \left| \omega_3 \right|$, leading to nuclear spin Rabi frequencies at $2 \left| \left| \omega_1 \right| \pm \left| \omega_2 \right| \pm \left| \omega_3 \right| \right|$. In general, $\left| \omega_i \right| \neq \left| \omega_j \right|$ for $i \neq j$ due to the different phases of the three $A_2^j$, so we predict that nuclear spin Rabi oscillations generically have four different frequencies of oscillation. This is qualitatively supported by the experimental data, which clearly shows oscillations of more than a single frequency.

We also note that the separate Rabi frequencies are highly dependent on the angle $\theta$ of the driving field, i.e. the relative angle between the RF magnetic field and the hBN crystal orientation. As an estimation, we take the averaged case 
and approximate $\left| A_1^j e^{i \theta_j} + \left(A_2^j\right)^* e^{-i \theta_j} \right| \sim \sqrt{\left| A_1^j \right|^2 + \left| A_2^j \right|^2} = \sqrt{\left| A_1 \right|^2 + \left| A_2 \right|^2}$ (as $\left| A_{1,2}^i \right| = \left| A_{1,2}^j \right|$ by symmetry). The nuclear spin Rabi frequencies then become $2 \gamma_e B_\mathrm{dr} \frac{\sqrt{\left| A_1 \right|^2 + \left| A_2 \right|^2}}{D_{gs} - \gamma_e B_z}$ (threefold degeneracy) and $6 \gamma_e B_\mathrm{dr} \frac{\sqrt{\left| A_1 \right|^2 + \left| A_2 \right|^2}}{D_{gs} - \gamma_e B_z}$, corresponding to a slow oscillation (in the general case, three slow oscillations) and a fast oscillation. We can approximately map the Rabi oscillation timescale to that of the slow oscillation, giving an effective nuclear gyromagnetic ratio of
\begin{equation} \label{nucapprox}
    \gamma_n^{\mathrm{eff}} = \frac{\Omega_n}{B_\mathrm{dr}}\approx 2 \gamma_e \frac{\sqrt{\left| A_1 \right|^2 + \left| A_2 \right|^2}}{D_{gs} - \gamma_e B_z} = \dfrac{\gamma_e \sqrt{A^2_{xx}+A^2_{yy}+2A^2_{xy}}}{\sqrt{2}(D_\mathrm{gs}-\gamma_e B_z)}
\end{equation}

\subsection{Experimental characterization of effective nuclear gyromagnetic ratio}\label{expermentalGamma_n}

\begin{figure}[t]
    \centering
    \includegraphics[width=0.65\textwidth]{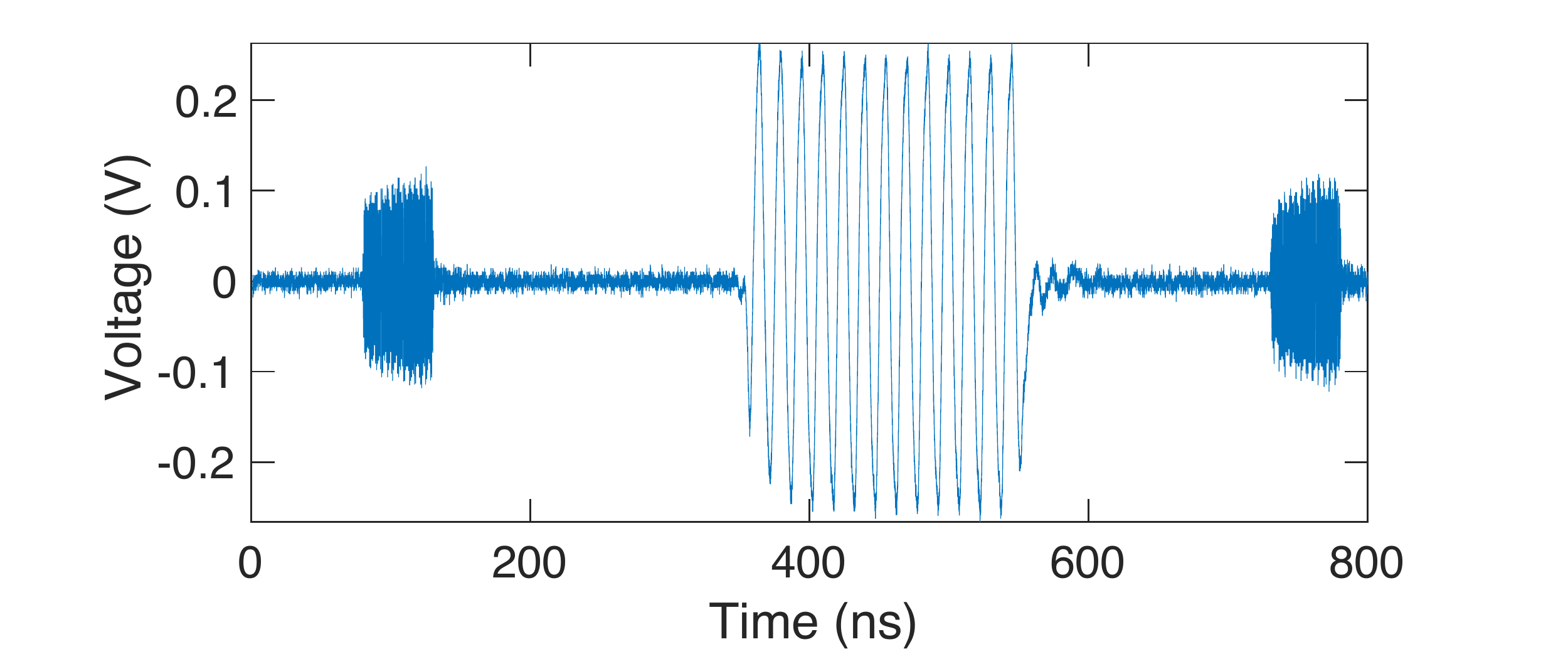}
    \caption{ {\bf Characterization of Microwave and RF power} Measured microwave and RF waveform after the stripline measured using an oscilloscope. Here we intentionally extend the microwave and RF pulse length to better quantify the amplitude ratios between the two drives. A 30~dB broadband RF and microwave attenuator is connected to the input of the oscilloscope for protection.}
    \label{Supp_Wav}
\end{figure}

To experimentally probe the effective gyromagnetic ratio of three nearest-neighbor \nfive nuclear spins, we first characterize the strength of the microwave pulse for electronic spin control and the RF pulse for nuclear spin control.
In particular, we directly measure the amplitude of the microwave and RF pulses by connecting the output from the coplanar waveguide to an oscilloscope with a sampling rate at $100~$GHz.
In Figure.~\ref{Supp_Wav}, we plot the measured waveform of the nuclear Rabi sequence, where we first apply a microwave $\pi$-pulse for electronic spin, a RF-pulse for nuclear spin, and then a final microwave $\pi$-pulse sequentially. 
By measuring the root-mean-square (RMS) voltage of each pulse, we obtain the amplitude ratio between RF and microwave fields
\begin{equation}
    R_\mathrm{volt} = \dfrac{V_\mathrm{RF}}{V_\mathrm{MW} }\approx 2.6.
\end{equation}
At these powers, the electronic and nuclear spin Rabi frequencies are measured to be $\Omega_e  \approx (2\pi)\times41.67$ MHz and $\Omega_n \approx (2\pi)\times 1.67$ MHz respectively.
%
Since the Rabi frequency should be proportional to the driving field amplitude, after accounting their amplitude difference, the effective Rabi frequency for nuclear spin at the same power as electronic spin is  $\Omega_n^\mathrm{eff}= \Omega_n/R_\mathrm{volt} \approx (2\pi)\times 0.64$ MHz.
This leads to an effective nuclear spin gyromagnetic ratio 
\begin{equation}
    \gamma^{\mathrm{eff}}_n = \frac{\Omega_n^\mathrm{eff}}{\Omega_e}\times\gamma_e\approx (2\pi)\times 0.043~\mathrm{MHz/G}
\end{equation}

Using our approximated nuclear gyromagnetic ratio from Eq.~\ref{nucapprox}, we find that 
\begin{equation}
    \dfrac{\gamma_n^\mathrm{eff}}{\gamma_e} \approx 99 \approx 2 \frac{\gamma_e}{\gamma_n} \frac{\sqrt{\left| A_1 \right|^2 + \left| A_2 \right|^2}}{D_{gs} - \gamma_e B_z} = \frac{\gamma_e \sqrt{A^2_{xx}+A^2_{yy}+2A^2_{xy}}}{\sqrt{2}(D_\mathrm{gs}-\gamma_e B_z)}.
\end{equation}
Using $\gamma_e/\gamma_n \approx 6487$ for \nfive nuclear spins and $D_\mathrm{gs}-\gamma_e B_z \approx (2\pi)\times 1.39~$GHz at esLAC, we estimate the transverse hyperfine interaction term to be $\sqrt{A^2_{xx}+A^2_{yy}+2A^2_{xy}} \approx (2\pi)\times 30~$MHz.
We note that this value turns out to be much smaller than the value predicted by previous ab-initio calculations, $\sqrt{A^2_{xx}+A^2_{yy}+2A^2_{xy}} \approx (2\pi)\times 142~$MHz \cite{ivady2020ab, gao2022nuclear}, by a factor of approximately 4.8.
%
%
In the next section, we present more experimental evidence on the nuclear spin resonance and Rabi oscillation measurement to support our observation.

\subsection{Simulation of Nuclear Spin Resonance and Rabi Oscillations}

In this subsection, we provide further experimental and theoretical evidence on the discrepancy between the measured transverse hyperfine interaction from our experiment and previous ab-initio calculations.
We begin by examining the nuclear spin resonance spectrum (main text Fig.~4b), as a large transverse hyperfine interaction will lead to visible shifts of each nuclear spin transition.
To simulate the experimental process, we first generate the Hamiltonian including the \vbm electronic spin ground state and the three nearest-neighbour \nfive nuclear spins (main text Eqn.~1 and Eqn.~\ref{eq14}). 
Here, we have included all hyperfine terms in the Hamiltonian.
After diagonalizing the Hamiltonian, we obtain the eigenenergies $E_n$ and the associated eigenstates $v_n$ of the system. 
The microwave and RF transitions from an initial state $v_i$ to a final state $v_f$ then has an energy $\Delta E= |E_f-E_i|$ and amplitude $ T_{i\rightarrow f} = |\langle v_f|S_x| v_i\rangle| + |\langle v_f|S_y| v_i\rangle|$.
We select the initial states $v_i$ within the manifolds $m_s = -1$ and $\sum m_I = 1/2$ as in the experiment and calculate the corresponding transition energies and amplitudes to all possible final states.
We focus on the frequency range of the nuclear spin transition ($\sim |A_{zz}| = (2\pi)\times 65.9~$MHz) and add a Lorentzian broadening with $2~$MHz FWHM to each transition to better reproduce the experimental results.

\begin{figure}
    \centering 
    \includegraphics[width = 0.7\textwidth]{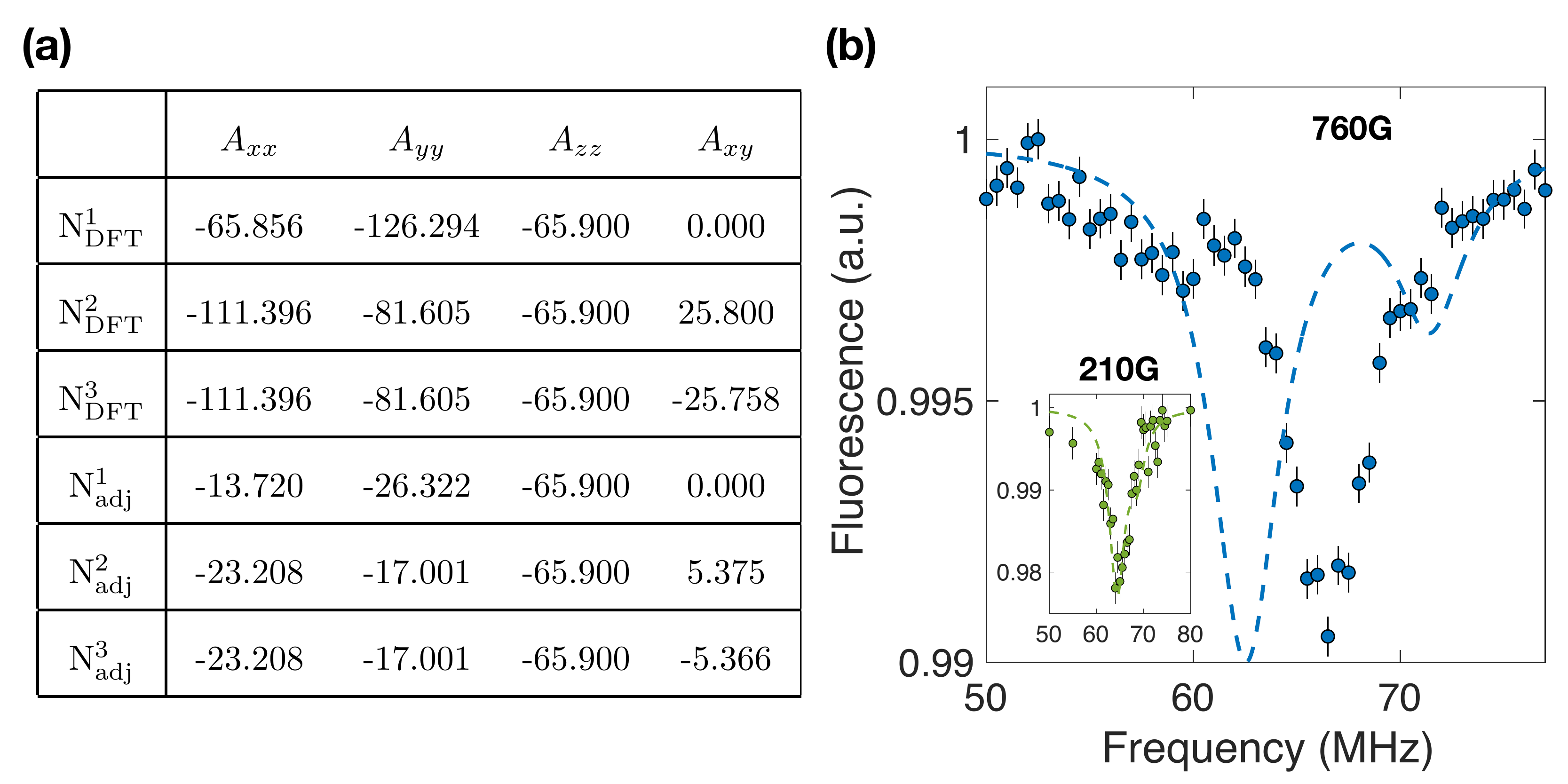}
    \caption{{\bf Nuclear Spin Resonance Measurement} (a) Hyperfine parameters $\mathbf{A}$ of the three nearest ${}^{15}$N nuclear spins. The first three rows are from ab-initio calculations \cite{ivady2020ab,gao2022nuclear}; the last three rows list the adjusted values by rescaling the transverse components ($A_{xx}$, $A_{yy}$, $A_{xy}$) by a factor of 4.8. (b) \nfive nuclear spin resonance spectra at $760~$G and $210~$G (Inset). Dashed lines show the numerical simulations using the ab-initio predicted hyperfine values.}
    \label{fig:Supp_Endor}
\end{figure}

If one directly uses the transverse hyperfine terms predicted by the ab-initio calculations, as shown in Figure~\ref{fig:Supp_Endor}, while the simulated nuclear spin spectrum still agrees with the experimental results at $B_z = 210~$G (Fig.~\ref{fig:Supp_Endor}b inset), it exhibits a significant shift at $B_z = 760~$G and cannot reproduce experimental results (Fig.~\ref{fig:Supp_Endor}b).
In comparison, we simply rescale the transverse hyperfine terms $A_{xx}$, $A_{yy}$, $A_{xy}$ values from ab-initio by a factor of 4.8 obtained by comparing theoretical and experimental Rabi oscillation timescales in the previous subsection. With these adjusted values, the simulated nuclear spin resonance spectra agree remarkably well with the experiments at at both $B_z = 210~$G and $B_z = 760~$G (main text Fig.~4c).
We note that, at small magnetic fields, the effect of transverse hyperfine coefficients is highly suppressed by the large splitting between \vbm $|0\rangle$ and $|-1\rangle$ spin states.
As a result, the simulation with the calculated and adjusted values fit the experimental data eqully well. 
At large external magnetic field, however, the electronic spin splitting between $|0\rangle$ and $|-1\rangle$ become much smaller, and the effect of transverse hyperfine coefficients becomes evident.

\begin{figure}[t]
    \centering
    \includegraphics[width=0.8\textwidth]{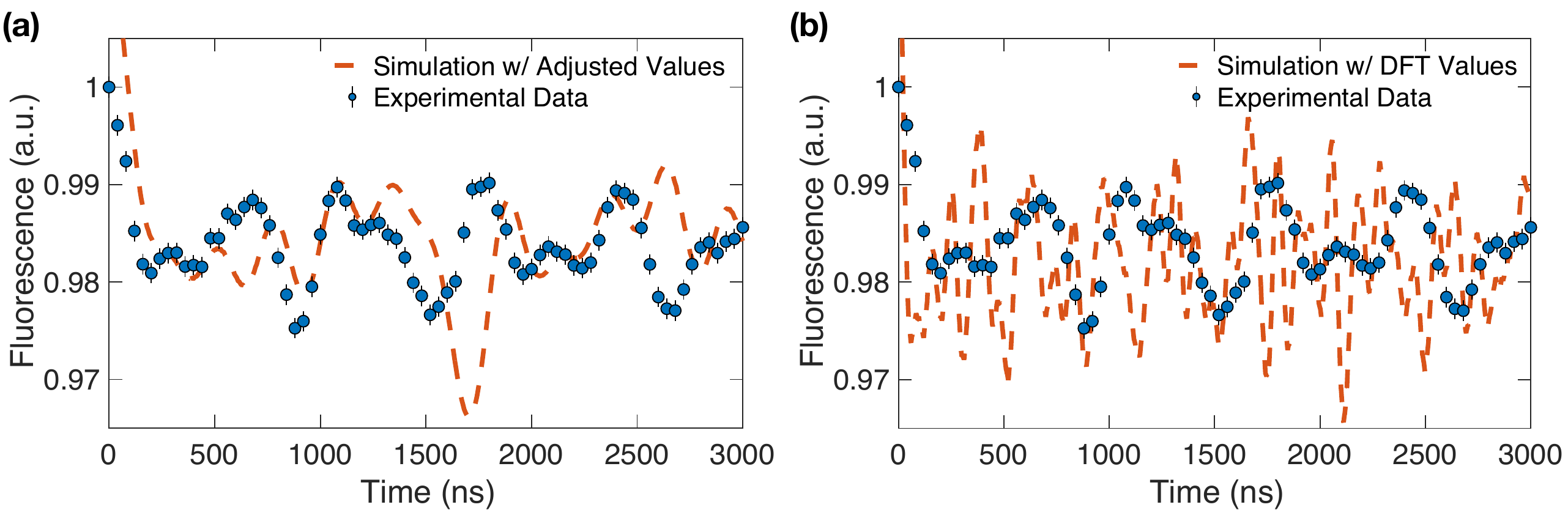}
    \caption{ {\bf Nuclear Spin Rabi Oscillations} Experimental data (blue points) superimposed with (a) a nuclear Rabi simulation using rescaled transverse hyperfine coupling strength and (b) the transverse hyperfine strength predicted by ab-initio calculations. The simulation using rescaled hyperfine parameters shows far better agreement with experiment but does not exactly match the trajectory due to unknown degrees of freedom in the angle $\theta$ (set to be 0 for simplicity) and the specific values of $A_1^j$ and $A_2^j$.}
    \label{Supp_NCRabi}
\end{figure}

We also perform simulations of the nuclear Rabi oscillation dynamics. 
Once again, simulations using transverse hyperfine values predicted by ab-initio yield significant disagreements from experiment, exhibiting Rabi oscillation signals at much faster oscillating frequencies than experimentally observed and indicating an exaggerated effective nuclear gyromagnetic ratio (Fig.~\ref{Supp_NCRabi}a).
On the other hand, simply rescaling the ab-initio values by a factor of 4.8 leads the simulated dynamics to display an oscillation with frequencies and patterns similar to the experiment, as shown in Figure~\ref{Supp_NCRabi}b. We do not see total agreement due to the existence of undetermined parameters such as the angle $\theta$ as well as the specific values of $A_1^j$ and $A_2^j$, which cannot be assumed to be simply rescaled from ab-initio calculations. Still, the theoretical and experimental dynamics show remarkable qualitative agreement.

\section{Optical Characterizations of \texorpdfstring{\vbm in hBN}{}}

\begin{figure}[ht]
    \centering
    \includegraphics[width=0.7\textwidth]{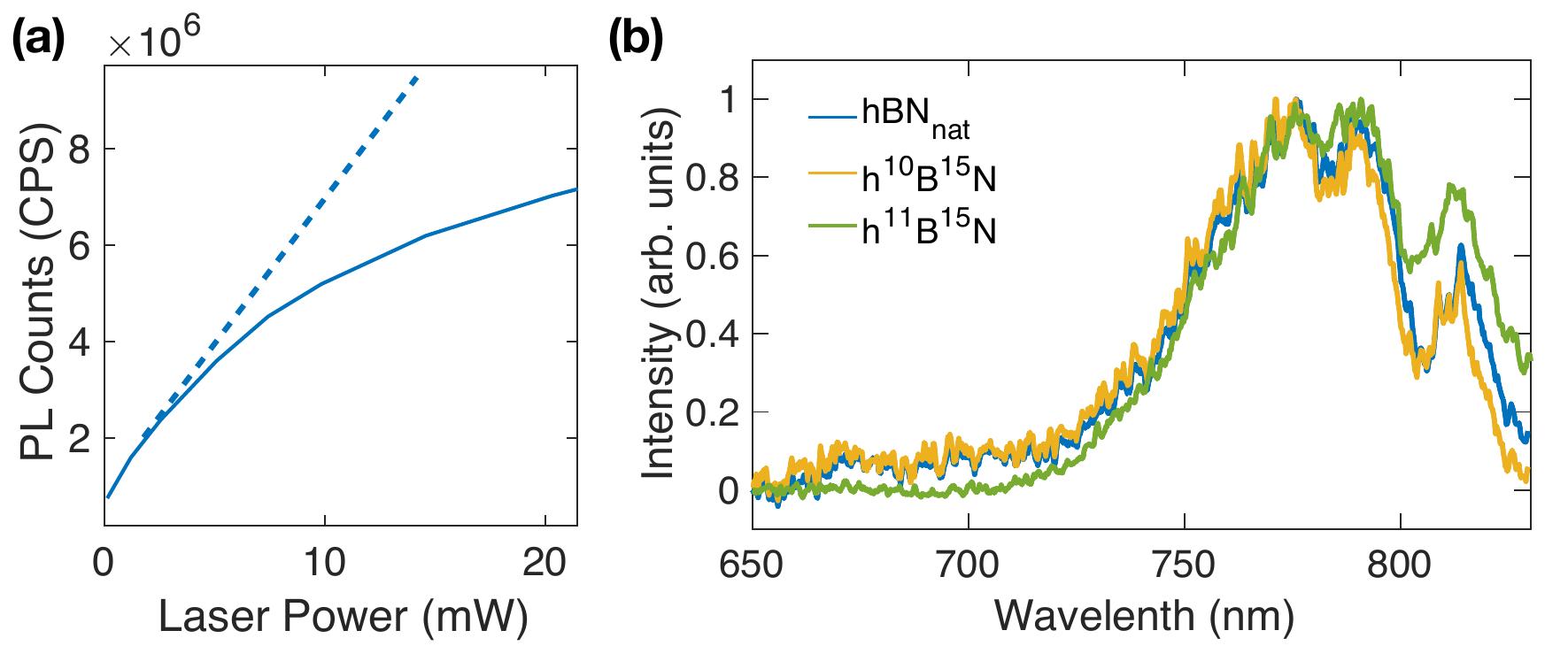}
    \caption{ {\bf Optical Characterization}
    (a) Optical saturation curve of \vbm in \hbn against laser power.
    (b) Photoluminescence spectrum of \vbm in \hbnnatns, \hbnns, and $\mathrm{h}{}^{11}\mathrm{B}{}^{15}\mathrm{N}$.
    }
    \label{Supp_OpticalC}
\end{figure}

To further characterize \vbm in \hbnns, we first perform an optical saturation measurement by monitoring the fluorescence while increasing laser power. 
As we can see from Figure \ref{Supp_OpticalC}a, the fluorescence increases with laser power and gradully deviates from a linear relation around $\sim 10~$ mW to saturation.
Note that this saturation point is also similar to that of the nuclear spin polarization (Fig.~3c from the main text).
This is not surprising since the nuclear spin polarize at esLAC through electronic spin and thus is also limited by the \vbm polarization level.
Additionally, we measure the photoluminescence spectrum of \vbm in \hbnnatns, \hbnns, and $\mathrm{h}{}^{11}\mathrm{B}{}^{15}\mathrm{N}$ and found no significant difference (Fig.~\ref{Supp_OpticalC}b).

\section{Extract sample thickness from the microscope optical image via color cycle fitting}\label{FindThickness}

In this section, we provide the method we use for estimating the sample thickness from the microscope optical image analysis, which is a common empirical technique in 2D materials to estimate the flake thickness~\cite{Puebla2022color}.
We first performed the optical imaging using a Zeiss AxioSkop optical microscope, with a fixed exposure time of 6 ms and automatic white balance adjustment.
The hBN flakes were exfoliated onto  Silicon wafers with 300 nm of thermal oxide, serving as a robust substrate.
To extract the wavelength of individual hBN flakes, $\lambda$, we first convert the picture RGB values to the HSV (Hue, Saturation, Value) color space and estimate $\lambda$ assuming a linear relationship between $\lambda$ and Hue value.

\begin{figure}
    \centering
    \includegraphics[width=0.9\textwidth]{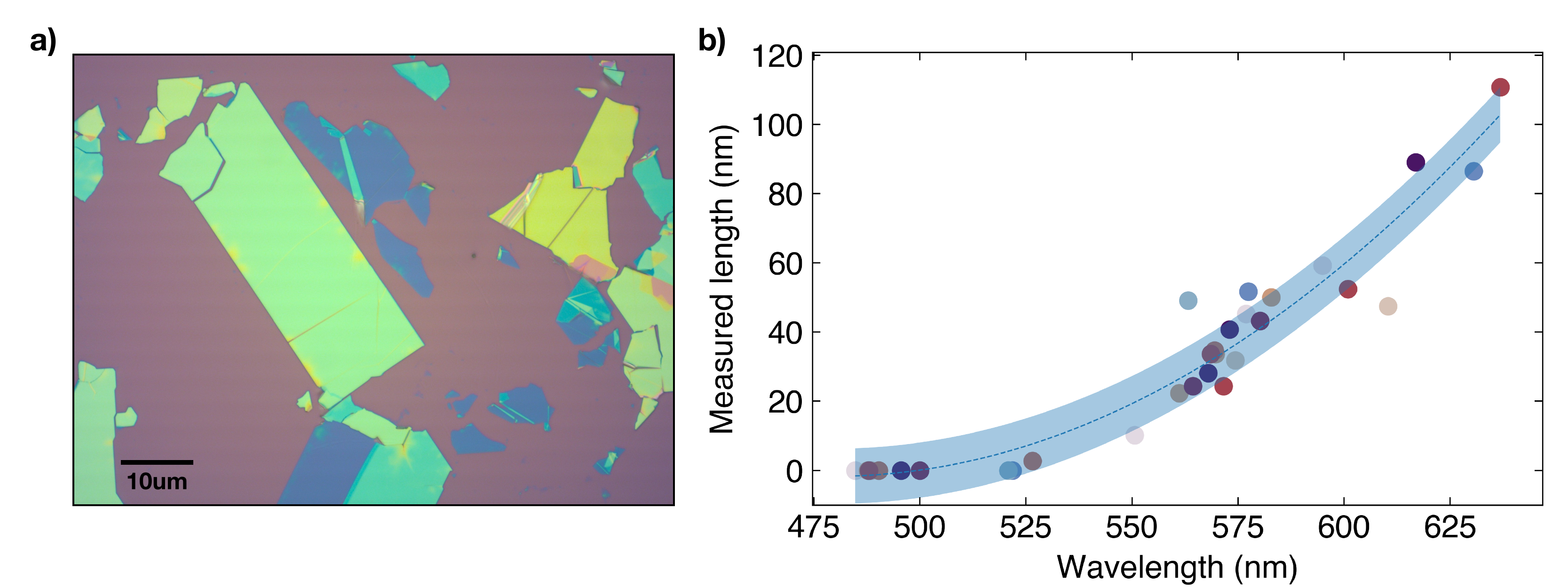}
    \caption{ {\bf Thickness calibration curve} (a) A common optical microscopy image of varying thickness hBN flakes on Si wafer with 300 nm thermal oxide. (b) Plotting of AFM thickness and wavelength estimation from optical image. The dashed line shows the second degree polynomial fitting. The shadowed region indicates the standard derivation of $\sigma$$\sim$8nm.
    The second degree polynomial fitting shown in the figure was then used for estimating the sample thickness.
    }
    \label{Supp_Thickness}
\end{figure}

In particular, we conducted AFM on $\sim30$ separate hBN flakes with varying thicknesses, establishing a reliable mapping from the color of the hBN flakes under our microscope to the precise thickness (Fig.~\ref{Supp_Thickness}). From the figure, we calibrate the thickness of the six flakes studied in our manuscript and summarize them in Supplementary Table \ref{tableR1}.

\begin{table}[h]
\begin{center}
\begin{tabular}{|m{10em}|m{5em} m{5em} m{5em} m{5em} m{5em} m{5em}|}
    \hline
    &S1&S2&S3&S4&S5&S6 \\ [1ex]
    \hline
    Thickness (nm) &64 $\pm$ 8 &63 $\pm$ 8 &63 $\pm$ 8 &72 $\pm$ 8 &70 $\pm$ 8 &21 $\pm$ 8 \\ [1ex]
    \hline
\end{tabular}
\caption{Summary of the estimated flake thickness for the six hBN samples S1-6.}\label{tableR1}
\end{center}
\end{table}

\bibliography{ref.bib}

\newpage

\newpage

\title{Methods: \texorpdfstring{\\}{} Isotope Engineering for Spin Defects in van der Waals Materials}

\author{
Ruotian~Gong,$^{1,*}$ 
Xinyi~Du,$^{1,*}$
Eli~Janzen, $^{2}$
Vincent~Liu,$^{3}$
Zhongyuan~Liu,$^{1}$
Guanghui~He,$^{1}$
Bingtian~Ye,$^{3}$ \\
Tongcang~Li,$^{4,5}$
Norman~Y.~Yao,$^{3}$
James~H.~Edgar,$^{2}$
Erik~A.~Henriksen,$^{1,6}$ 
Chong~Zu$^{1,6,\dag}$
\\
\medskip
\normalsize{$^{1}$Department of Physics, Washington University, St. Louis, MO 63130, USA}\\
\normalsize{$^{2}$Tim Taylor Department of Chemical Engineering,
Kansas State University, Manhattan, KS 66506, USA}\\
\normalsize{$^{3}$Department of Physics, Harvard University, Cambridge, MA 02138, USA}\\
\normalsize{$^{4}$Department of Physics and Astronomy, Purdue University, West Lafayette, Indiana 47907, USA}\\
\normalsize{$^{5}$Elmore Family School of Electrical and Computer Engineering, Purdue University, West Lafayette, IN 47907, USA}\\
\normalsize{$^{6}$Institute of Materials Science and Engineering, Washington University, St. Louis, MO 63130, USA}\\
\normalsize{$^*$These authors contributed equally to this work}\\
\normalsize{$^\dag$To whom correspondence should be addressed; E-mail: zu@wustl.edu}\\
}

\date{\today}

\maketitle

\section{hBN Device Fabrication}

\subsection{\texorpdfstring{Growth of Isotopically Purified \hbnns}{}}
High-quality \hbn flakes are grown via precipitation from a molten Ni-Cr flux as detailed in \cite{janzen2023boron}. 
Ni, Cr, and isotopically pure \bten powders in the mass ratio 12:12:1 [Ni:Cr:B] are first loaded into an alumina crucible, then heated at 1550°C for 24 hours in a continuously flowing Ar/$\mathrm{H}_2$ atmosphere to melt all three components into a homogenous ingot and remove oxygen impurities. 
Next, this ingot is heated at 1550 °C for 24 hours in a static ${}^{15}\mathrm{N}_{2}$/$\mathrm{H}_2$ atmosphere to saturate the molten flux with ${}^{15}\mathrm{N}$, then slowly cooled at 1°C/hr to precipitate hBN from the solution. 
Since isotopically pure \bten ($>99.2\%$) and \nfive ($>99.7\%$) are the only sources of boron and nitrogen in the system, the hBN single crystals that precipitated have the same isotopes.

\begin{figure}[b!]
    \centering
    \includegraphics[width=1\textwidth]{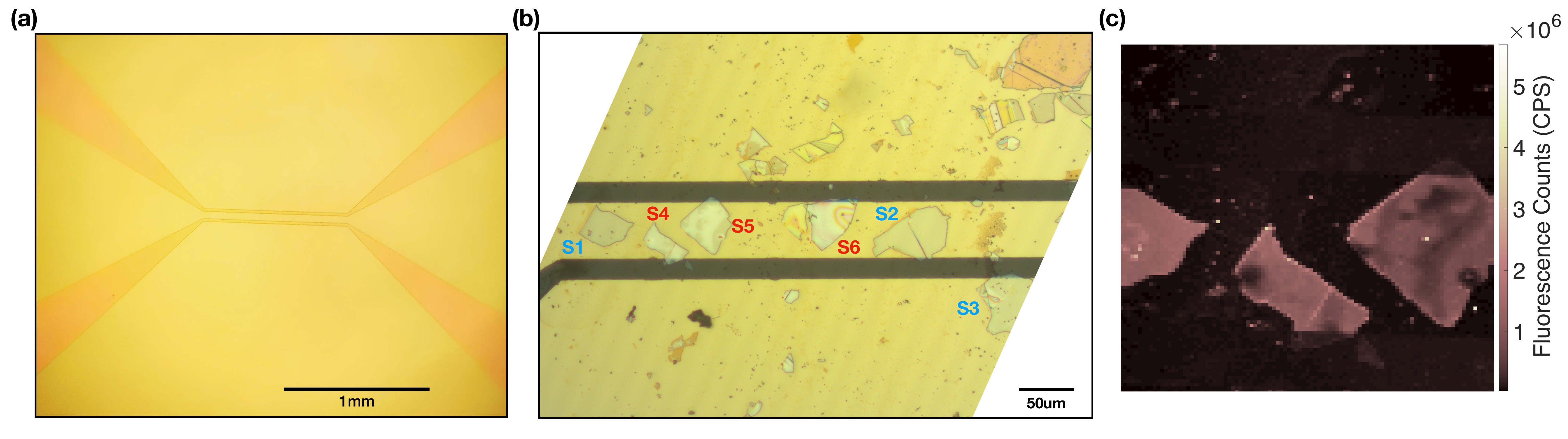}
    \caption{{\bf Optical image of the coplanar waveguide}
    (a) Overview of the coplanar waveguide before transferring the hBN flakes.
    (b) Zoomed-in microscope image of the coplanar waveguide after transferring the hBN samples. The six flakes studied in this work are labeled as S1-S6.
    (c) Fluorescence confocal scan of sample S1, S4, and S5.
    }
    \label{Supp_Stripline}
\end{figure}

\subsection{\texorpdfstring{Creation of \vbm}{} defects}
We first exfoliate hBN nanosheets using tapes and then transfer them onto the Si/SiO$_2$ wafer. 
The wafer is pretreated with O$_2$ plasma at $50~$W for 1 minute (flow rate $20~$sccm).
The tapes and wafers are heated to $100^\circ$C for 1 minute before samples are carefully peeled off for maximum hBN flake yield \cite{huang2015reliable}. 
The wafers with \hbnnat and \hbn are then sent to CuttingEdge Ions LLC for $^4$He$^+$ ion implantation with energy $3~$keV and dose density $1~\mathrm{ion}/\mathrm{nm}^{2}$ to create \vbm defects with estimated density $\sim 150~$ppm \cite{gong2023coherent}.
After implantation, the hBN flakes with thicknesses ranging from $20-70~$nm are transferred onto the coplanar waveguide using polycarbonate (PC) stamps.
Specifically, the temperature is raised in a stepwise manner to facilitate the transfer of hBN flakes. After successful transfer, the waveguide is immersed in chloroform for cleaning, removing any residual PC films.
For a systematic comparison between \vbm in isotopically purified \hbn and naturally abundant \hbnnatns, we transfer three \hbn flakes (S1-S3) and three \hbnnat flakes (S4-S6) onto a single waveguide for measurement (Figure~\ref{Supp_Stripline}).

\subsection{Fabrication of Coplanar Waveguide}
An impedance-matched ($50~\Omega$) coplanar waveguide is fabricated onto a $400~\mu$m thick sapphire wafer for microwave delivery (Figure~\ref{Supp_Stripline}).
Specifically, photoresist layers (LOR 1A/S1805) are first applied to the sapphire wafer at a spin rate $4000~$rpm for 1 minute.
A carefully designed waveguide pattern with a $50~\mu$m wide central line is then developed using a direct-write lithography system (Heidelburg DWL66+), followed by an O$_2$ plasma cleaning process to remove resist residue. 
A $10~$nm chromium adhesion layer and a $160~$nm gold layer are deposited using thermal evaporation, followed by a lift-off process.



\section{Comparison of \texorpdfstring{\vbm}{} in hBN with different isotopes}

\subsection{Simulation of ESR spectrum}
\begin{figure}[b]
    \centering
    \includegraphics[width=0.55\textwidth]{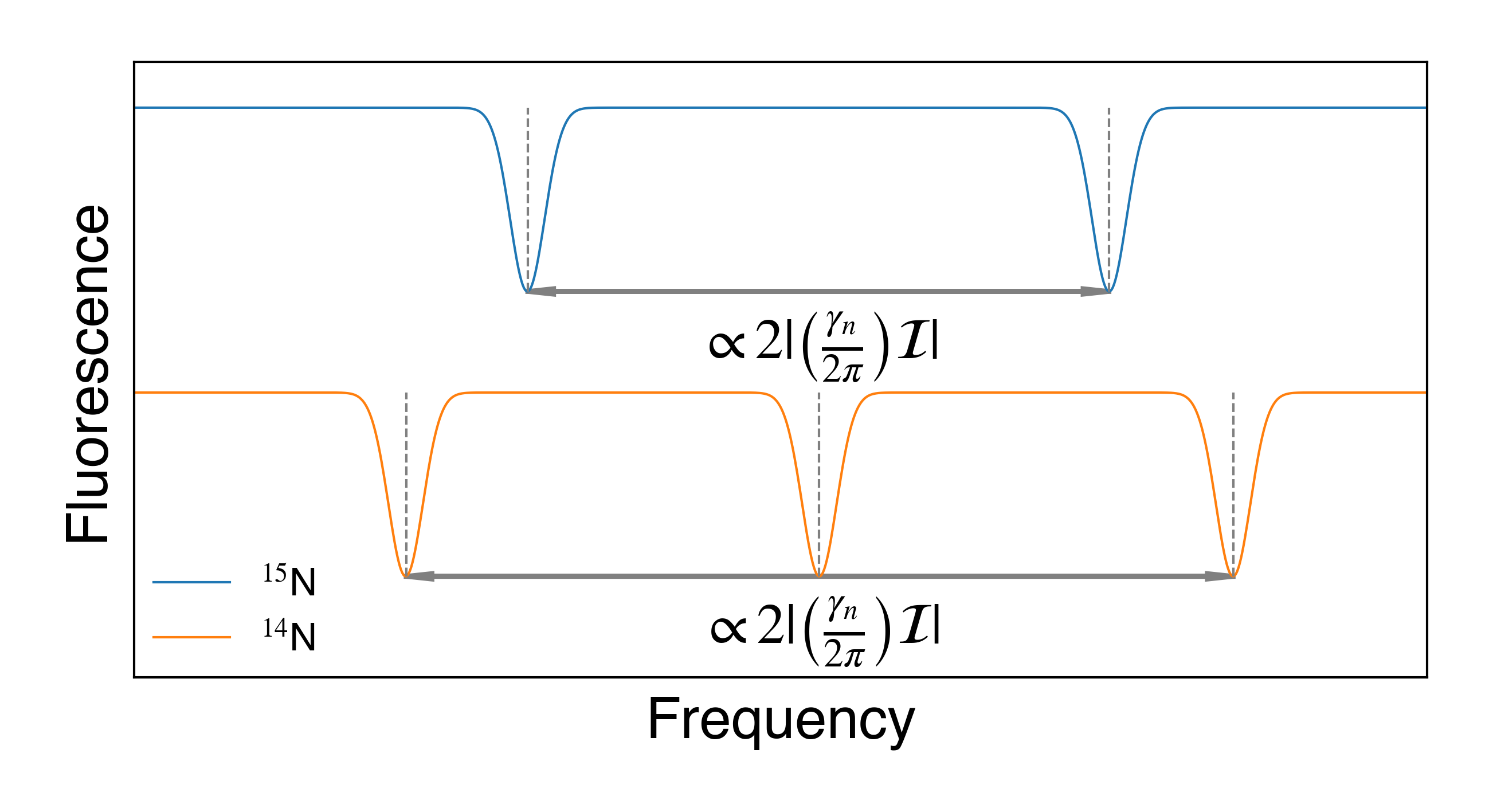}
    \caption{{\bf Schematic ESR Spectrum} A schematic ESR spectrum of \vbm coupling to a single \nfour (orange) or \nfive (blue) nuclear spin with a overall bandwidth of the transitions proportional to $2|(\frac{\gamma_n}{2\pi})\mathcal{I}|$.
    }
    \label{Supp_2gammanI}
\end{figure}

To quantitatively capture the measured ESR spectra in hBN samples with different isotopes, we numerically simulate the \vbm transitions by accounting for its hyperfine couplings to the nearest $36$ nuclear spins. 
At small external magnetic fields, the hyperfine terms from main text Eqn.~(1), can be approximated as  $\sum_{j}\mathbf{S}\mathbf{A}^j\mathbf{I}^j \approx\sum_{j}A_{zz}^j S_z I_z^j = (\sum_{j} A_{zz}^j I_z^j) S_z$, which leads to an effective energy shift of the electronic spin state depending on the nearby nuclear spin configurations.
Considering the hyperfine interaction between \vbm and an individual nuclear spin with spin number $\mathcal{I}$, the averaged spectrum for the electronic spin transition $|m_s=0\rangle \leftrightarrow |m_s=\pm1\rangle$ will exhibit $2\mathcal{I}+1$ resonances with splitting $A_{zz}$ assuming the nuclear spin is in a fully mixed state (to be expected at room temperature). 
We note that the value of the hyperfine coupling is proportional to the gyromagnetic ratio of the nuclear spin, i.e. $A_{zz} \propto\gamma_n$; therefore, the overall bandwidth of the transition is proportional to $2\mathcal{I}\times |\gamma_n|$.
Figure~\ref{Supp_2gammanI} shows a schematic ESR spectrum assuming the \vbm couples to a single \nfour or \nfive nuclear spin.
To achieve a smaller overall linewidth of the \vbm transitions, the value 2$|\gamma_n \mathcal{I}|$ must be minimized for each atomic species in the host material (main text Figure.~1a).

%

\begin{figure}[t!]
    \centering
    \includegraphics[width=0.8\textwidth]{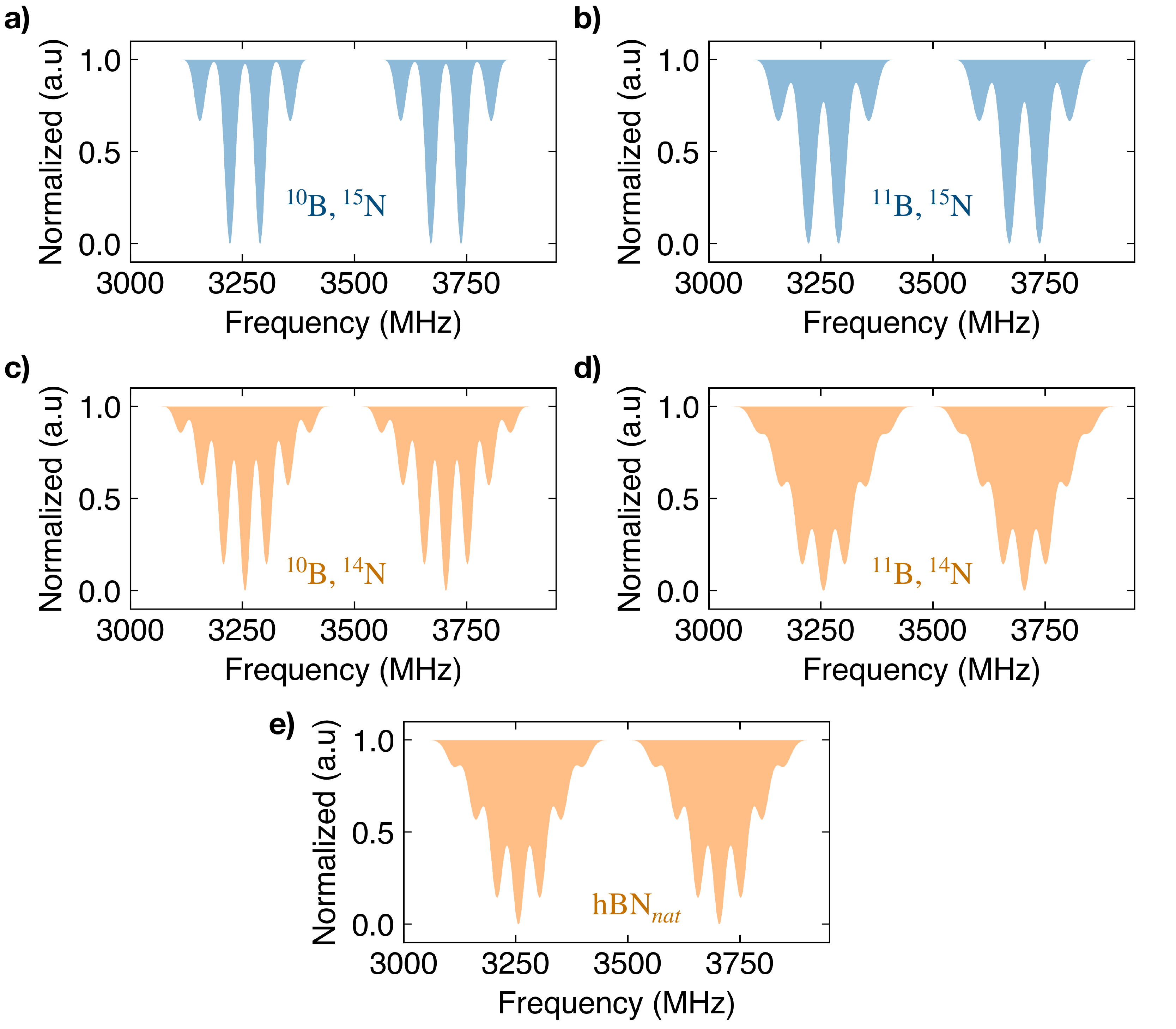}
    \caption{{\bf Simulated ESR Spectra for Different Isotopes} Simulated ESR spectra of \vbm in the four different possible isotopically purified hBN crystals (a) $\mathrm{h}{}^{10}\mathrm{B}{}^{15}\mathrm{N}$, (b) $\mathrm{h}{}^{11}\mathrm{B}{}^{15}\mathrm{N}$, (c) $\mathrm{h}{}^{10}\mathrm{B}{}^{14}\mathrm{N}$, (d) $\mathrm{h}{}^{11}\mathrm{B}{}^{14}\mathrm{N}$, and (e) in \hbnnatns.}
    \label{Supp_5isotopes}
\end{figure}

Following this approximation, we proceed to compute the \vbm ESR spectrum by summing over the hyperfine couplings to the 18 nearest nitrogen and 18 nearest boron nuclear spins. The values of $A_{zz}$ for \nfour and \belev have been previously predicted using a Density Functional Theory (DFT) calculation \cite{ivady2020ab}.
To calculate $A_{zz}$ for \nfive and \bten nuclear spins, we only need to account for the differences between gyromagnetic ratios.
Direct summation of all nuclear configurations yields $(2 \mathcal{I}_N +1)^{18} \times (2 \mathcal{I}_B + 1)^{18}$ individual resonances, where $\mathcal{I}_N$ and $\mathcal{I}_B$ represent the nuclear spin numbers of the corresponding nitrogen and boron isotopes, respectively. 
To mitigate the computational cost, we employ Fourier transformation to simplify the calculations (see Supplementary Note 2).
The final ESR spectrum is obtained by binning transition frequencies with bin size $0.5~$MHz.
Figure~\ref{Supp_5isotopes} summarizes the simulated ESR spectra of \vbm centers in the four different types of isotopically purified hBN crystals ($\mathrm{h}{}^{10}\mathrm{B}{}^{14}\mathrm{N}$, $\mathrm{h}{}^{10}\mathrm{B}{}^{15}\mathrm{N}$, $\mathrm{h}{}^{11}\mathrm{B}{}^{14}\mathrm{N}$, $\mathrm{h}{}^{11}\mathrm{B}{}^{15}\mathrm{N}$), as well as naturally abundant hBN.
As predicted, \vbm centers in \hbn has the least crowded spectrum with the narrowest individual transitions.

\subsection{Coherence measurement of \texorpdfstring{\vbm in other isotopically purified hBN}{}}
\begin{figure}[ht]
    \centering
    \includegraphics[width=0.6\textwidth]{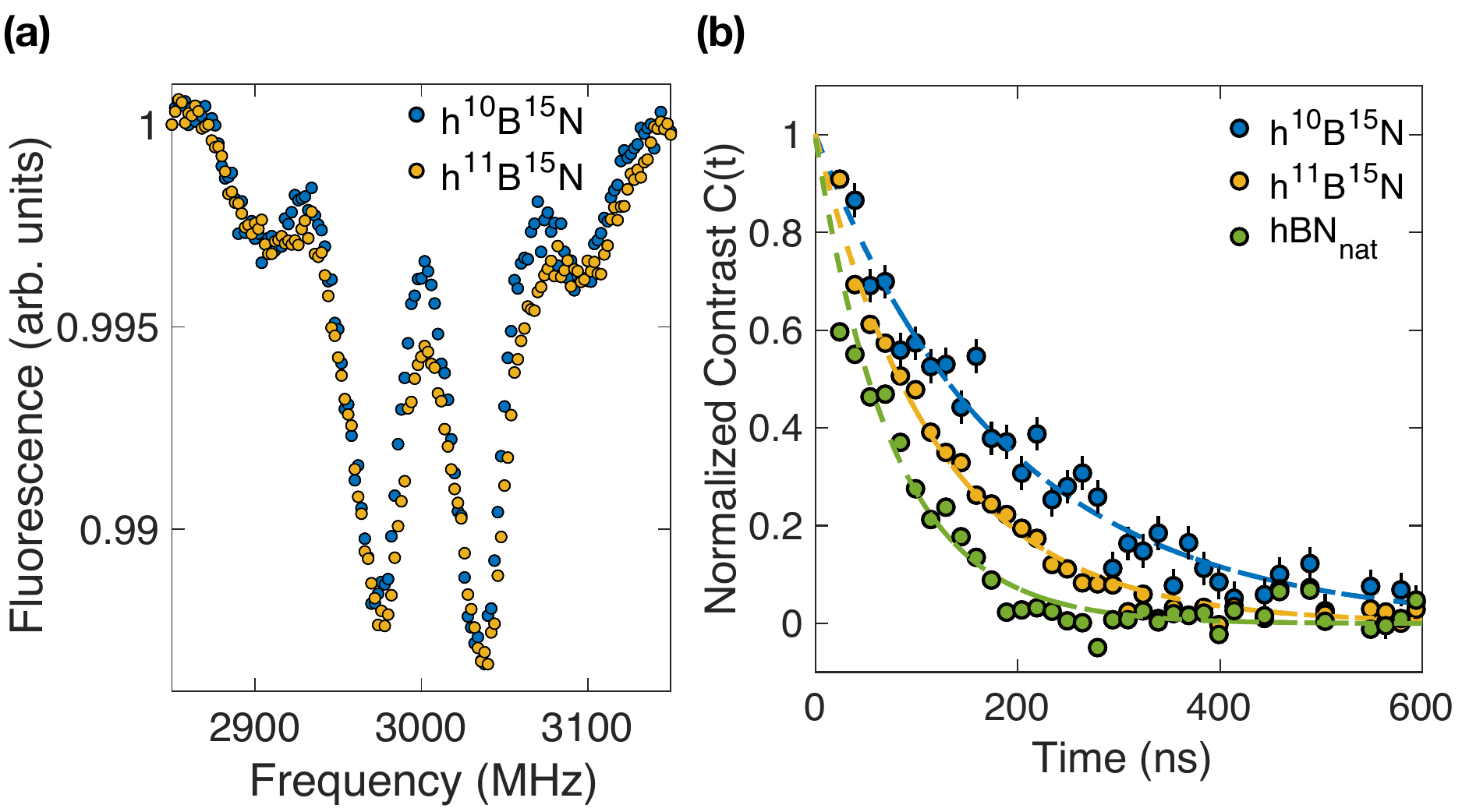}
    \caption{{\bf \hbnelev characterization}.
    (a) The ESR spectra of \vbm in \hbn and \hbnelevns.
    (b) The spin echo measurement. $T_2^\mathrm{E} \approx 186 \pm 22,~ 119 \pm 6,~ 76 \pm 7~$ns for \hbnns, \hbnelevns, and \hbnnat respectively. Error bars represent 1 s.d. accounting statistical uncertainties. Source data are provided as a Source Data file.
    }
    \label{Supp_B11}
\end{figure}

To further verify that \hbn is the ideal host regarding the \vbm coherence timescales, we prepare another sample with $\mathrm{h}{}^{11}\mathrm{B}{}^{15}\mathrm{N}$, which is predicted to be the second best host material. 
Unsurprisingly, the ESR spectrum of \vbm in \hbnelev exhibits a slightly broader transition linewidth in comparison with \hbn (Fig.~\ref{Supp_B11}a).
By measuring the Spin Echo coherence time, \vbm in \hbnelev indeed shows $T_2^\mathrm{E} = 119\pm6~$ns, in between the values measured from \hbn and \hbn (Fig.~\ref{Supp_B11}b). 
A previous work \cite{haykal2022decoherence} has also shown that \vbm in $\mathrm{h}{}^{10}\mathrm{B}{}^{14}\mathrm{N}$ and $\mathrm{h}{}^{11}\mathrm{B}{}^{14}\mathrm{N}$ did not exhibit a clear improvement in spin echo $T_2$ compared to natural abundant hBN.
Therefore, we believe that \hbn is indeed the optimal isotope selection for improving the coherence time of \vbmns.

\section{Quantifying nuclear polarization}

To quantify the nuclear spin polarization of the nearest three \nfive, we assume each nuclear spin is individually prepared to $|m_I=\:\uparrow \rangle$ with probability $P(\ket{\uparrow})$ and $|m_I=\:\downarrow \rangle$ with probability $1-P(\ket{\uparrow})$. 
%
The energy sublevels accounting for the three \nfive spins should effectively follow a bionomial distribution with $n=3$, the total number of a sequence of independent events.
For a random variable $X$ defined as the number of \nfive nuclear spins being measured to be $|m_I=\:\uparrow \rangle$ in a single observation, the probability mass function is
\begin{equation} \label{eq5}
    \mathrm{Pr} (X = k) = \binom{n}{k}P^k(1-P)^{3-k}.
\end{equation}
For example, the four possible nuclear spin configurations correspond to $\sum m_I = {-\frac{3}{2}, -\frac{1}{2}, \frac{1}{2}, \frac{3}{2}}$ and $X = {0, 1, 2, 3}$.
Therefore, the theoretical probability distribution follows $\{\binom{3}{3} (1-P)^3 : \binom{3}{2} P(1-P)^2 : \binom{3}{1} P^2(1-P) : \binom{3}{0} P^3 \} = \{(1-P)^3 :3 P(1-P)^2 : 3 P^2(1-P) : P^3\}$. 
When there is no nuclear spin polarization ($P=0.5$), we obtain amplitudes $\{1:3:3:1\}$ for the ESR spectra.

To quantitatively extract the nuclear polarization, we first fit the ESR spectrum using the sum of four equally spaced Lorentzian distribution whose amplitude follows the aforementioned ratio.
However, we note that the gyromagnetic ratio for \nfive is negative, which reverses the order of the sublevels in the ESR spectrum. 
In this case, for the \vbm transition between electronic $|m_s=0\rangle$ and $|m_s = +1\rangle$ states, the lowest-frequency resonance corresponds to $\sum m_I = \frac{3}{2}$ or $X=3$ (main text Figure~3b).

\begin{figure}
    \centering 
    \includegraphics[width = 0.5\textwidth]{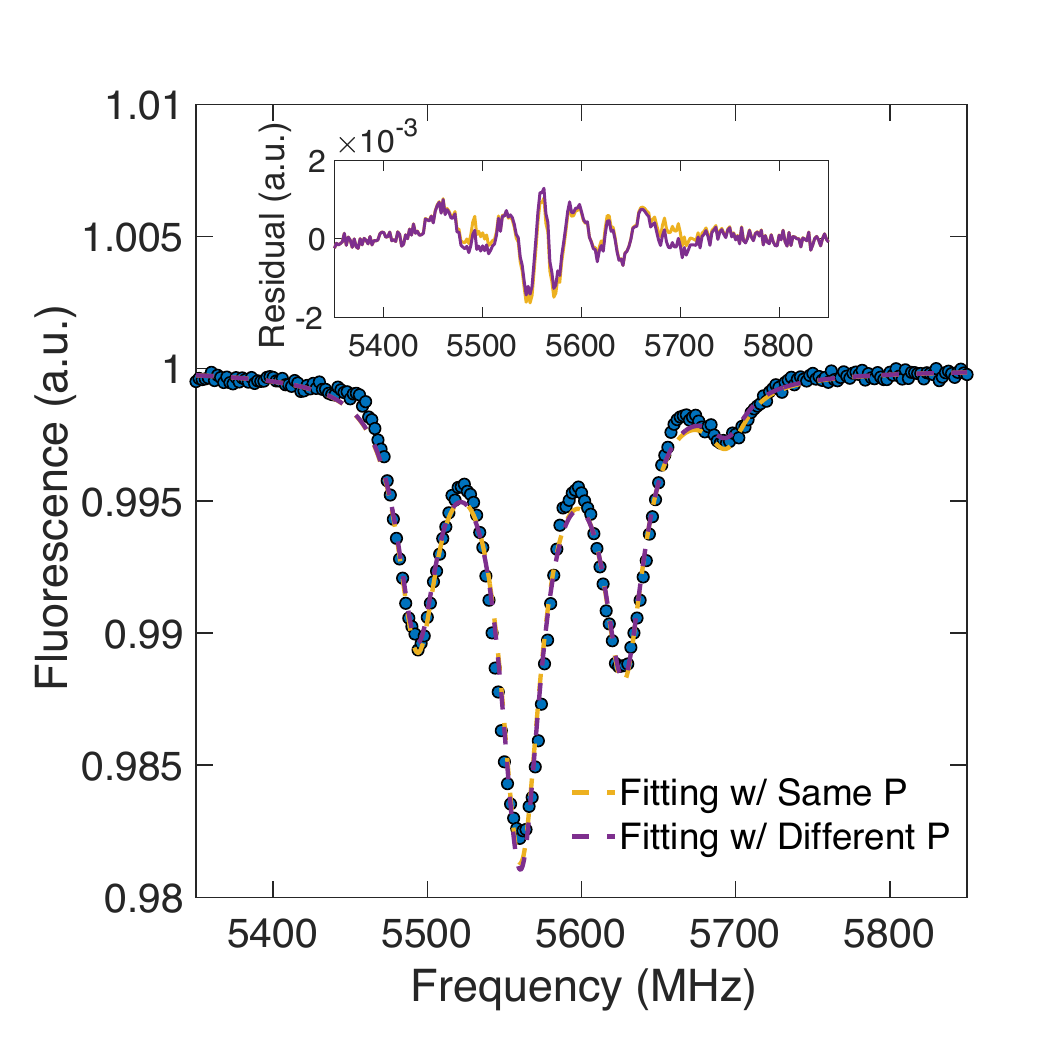}
    \caption{{\bf Extracting Nuclear Spin Polarization} Fits of the ESR spectrum at esLAC using two different methods: assuming independently polarization probability $P$ for all three \nfive nuclear spins, yielding the amplitude ratio $\{(1-P)^3 :3 P(1-P)^2 : 3 P^2(1-P) : P^3\}$ (yellow dashed lines) and assuming two different nuclear polarization probabilities, $P_1$ and $P_2$, yielding amplitude ratio $\{(1-P_1)^2(1-P_2): (1-P_1)^2P_2+2P_1(1-P_1)(1-P_2):P_1^2(1-P_2)+2P_1P_2(1-P_1): P_1^2P_2\}$ (purple dashed lines).
    Inset: computed residuals between experimental data and two fits.}
    \label{fig:Supp_Pola}
\end{figure}

We note here that previous studies have predicted a reduction of symmetry in the \vbm excited state manifold from $\mathrm{D}_\mathrm{3h}$ to $\mathrm{C}_\mathrm{2v}$ due to Jahn-Teller effect \cite{reimers2020photoluminescence, ivady2020ab, gao2022nuclear}.
This can lead to a different hyperfine interaction for one of the \nfive nuclear spins compared to the other two. 
To this end, we also investigate the fitting of the experimental spectra assuming two of the \nfive nuclear spins are polarized to $|m_I=\:\uparrow \rangle$ with probability $P_1$ and the third \nfive is polarized with probability $P_2$.
The corresponding probability distribution for the four hyperfine resonances takes the form $ \{(1-P_1)^2(1-P_2): (1-P_1)^2P_2+2P_1(1-P_1)(1-P_2):P_1^2(1-P_2)+2P_1P_2(1-P_1): P_1^2P_2\}$.
Figure~\ref{fig:Supp_Pola} illustrates a comparison between the two different fitting methods, and we find that both models capture the experimental spectra almost identically well. Thus, the addition of the parameter $P_2$ does not significantly enhance agreement with experiment.
Furthermore, the two-parameter fit has much greater uncertainty in the extracted values of $P_1$ and $P_2$ compared to the single-parameter fit and the net nuclear spin polarization amount remains almost exactly the same between the two cases.
Therefore, we utilize the first model with minimal fit parameters throughout the manuscript.
Further studies are required to experimentally distinguish whether a parameter such as $P_2$ is required.


\bibliography{ref.bib}